\documentclass[12pt]{JHEP3}

\usepackage{amsmath}

\newcommand{\e}{\epsilon}

\renewcommand{\L}{{\mathcal{L}}}
\newcommand{\bL}{\bar{{\mathcal{L}}}}
\newcommand{\z}{{\bar z}}

\renewcommand\O{{\mathcal{O}}}
\newcommand{\be}[1]{ \begin{equation}\label{#1} }
\newcommand{\ee}{\end{equation}}
\newcommand{\ben}[1]{\begin{eqnarray}\label{#1} }
\newcommand{\een}{\end{eqnarray}}
\newcommand{\eq}[1]{(\ref{#1})}
\newcommand{\p}{\partial}
\newcommand{\D}{\Delta}
\newcommand{\refb}[1]{(\ref{#1})}

\title{GCA in 2d}

\author{
Arjun Bagchi$^1$, Rajesh Gopakumar$^1$, Ipsita Mandal$^{1,2}$, Akitsugu Miwa$^1$\\
$\;$ $^1$Harish-Chandra Research Institute,\\
$\;$ $\,$Chhatnag Road, Jhusi,\\
$\;$ $\,$Allahabad 211019, India\\

$\;$ $^2$LPTHE, Universite Pierre et Marie Curie,\\
$\;$ $\,$Paris 6, 4 Place Jussieu,\\
$\;$ $\,$75252 Paris Cedex 05, France\\

$\;$\email{arjun, gopakumr, ipsita, akitsugu@hri.res.in}
}

\abstract{We make a detailed study of the infinite dimensional 
Galilean Conformal Algebra (GCA) in the case of 
two spacetime dimensions. Classically, this algebra is precisely obtained from a contraction of the 
generators of the relativistic conformal symmetry in $2d$. Here we find quantum mechanical 
realisations of the (centrally extended) GCA by considering scaling limits of certain 2d CFTs. 
These parent CFTs are non-unitary  and have their left and right central charges become large in magnitude and opposite in sign. We therefore develop, in parallel to the usual machinery for $2d$ CFT, many of the tools for the analysis of the quantum mechanical GCA. 
These include the representation theory based on GCA primaries, Ward identities for their correlation functions and a nonrelativistic Kac table. In particular, the null vectors  of the GCA lead to differential equations for the four point function. 
The solution to these equations in the simplest 
case is explicitly obtained and checked to be consistent with various requirements. 
}

\preprint{HRI/ST/0923}

\begin{document}

\baselineskip 3.5ex

\section{Introduction}

\subsection{Non-relativistic Conformal Symmetries}

Conformal symmetry has gradually moved from being a mere mathematical curiosity (e.g. as an invariance of certain classical equations of motion like the Maxwell and more generally Yang-Mills equations)  to playing a central dynamical role governing the quantum behaviour of field theories through RG fixed points. Thus an understanding of CFTs in various dimensions has had a major impact on physics, starting with the study of critical phenomena all the way through to the more recent AdS/CFT correspondence. 

In all this, the focus has largely been on relativistic conformal field theories. Non relativistic versions of conformal symmetry have been somewhat studied in the context of the so-called Schrodinger symmetry 
\cite{Hagen:1972pd, Niederer:1972zz, Henkel:1993sg, Nishida:2007pj}
which is an enhanced symmetry that arises in taking the nonrelativistic limit of the massive Klein-Gordon equation. In $d$ spacetime dimensions, this group has, in addition to the ${d(d+1)\over 2}$ parameters of the Galilean group, a nonrelativistic dilatation $x_i \rightarrow \lambda x_i, t\rightarrow \lambda^2 t$
and {\it one} more conformal generator. 
Much less explored are the nonrelativistic limits of the full $SO(d,2)$ conformal invariance. 
In fact, a non-relativistic group contraction of $SO(d,2)$ gives rise to a ${(d+1)(d+2)\over 2}$ parameter group (see for e.g. \cite{negro1997, Lukierski:2005xy} and in an $AdS$ context \cite{Gomis:2005pg}). This contains the Galilean symmetries together with a uniform dilatation  $x_i \rightarrow \lambda x_i, t\rightarrow \lambda t$ and  $d$ other generators which are the analogues of the special conformal transformations. 

Quite remarkably, the actual set of conformal isometries of nonrelativistic spacetime is much larger than the above contracted version of $SO(d,2)$ \cite{Bagchi:2009my}. In {\it every} spacetime dimension it is actually an infinite dimensional algebra which was dubbed the Galilean Conformal Algebra (GCA). This algebra consists of local conformal transformations acting on time (generating a single copy of the Virasoro algebra) together with a current algebra for rotations as well as arbitrary time dependent boosts in the spatial directions. It was shown in \cite{Duval:2009vt} that this infinite dimensional algebra can also be obtained by considering the natural nonrelativistic limit of the relativistic conformal killing equations and is the maximal set of such nonrelativistic conformal isometries.\footnote{For related work on various aspects of the GCA, see \cite{Duval:2009vt}--\cite{IPM}.} The situation is very analogous to that in the relativistic $d=2$ case where one has two copies of the Virasoro algebra generating the maximal set of local conformal isometries.

\subsection{The role of the GCA in physical systems}

At this point it is worthwhile to remind the reader why one should be intersted in the GCA. In \cite{Bagchi:2009my}, the motivation for looking at the contracted algebra was to construct a systematic non-relativistic limit of the AdS/CFT conjecture\cite{Maldacena:1997re}. The non-relativistic AdS/CFT correspondence is one of the avenues of exploration of the recent studies of condensed matter systems using the gauge-gravity duality (for a review and a list of relevant papers, see \cite{Hartnoll}). By taking a parametric limit of the conformal algebra in \cite{Bagchi:2009my}, we wished to isolate a sector of AdS/CFT in the spirit of the BMN limit\cite{BMN}. Though the realization of the bulk symmetry was very different from BMN, there were novel structures discovered on both sides of the duality. We hope that further explorations of the GCA would lead to a better understanding of the AdS/CFT correspondence which even today is not properly understood.

On the other hand, the GCA may play a role in real life physical systems. 
It was shown in \cite{Bhattacharyya:2008kq} (see also \cite{Fouxon:2008tb, Fouxon:2008ik}) that the non-relativistic Navier-Stokes equation has an invariance under the generators of boosts and constant accelerations. In fact, the algebra of invariances found there is precisely the algebra of the finite GCA (with dilatations also being an invariance when the viscous term is dropped) \cite{Bagchi:2009my}. 
In fact, all the time dependent boosts {\it are} invariances of the local equations (both the Navier-Stokes and the Euler equations)\cite{Russian}{\footnote{This is so despite the interpretational claim to the contrary in \cite{Horvathy:2009kz}.}}. The Navier-Stokes equation is believed to be realized in the non-relativistic limit of the hydrodynamic regime of all quantum field theories. The realization of a part of the infinite GCA as its symmetries possibly indicates a universality of the existence of the GCA. 

Our focus in this paper would be on the two dimensional GCA. It is worthwhile to mention that a mathematically identical infinite dimensional algebra (called ${\rm alt}_1$) in the case of one spatial dimension had also appeared independently in \cite{Henkel06} in the context of statistical mechanics {\footnote{It was also noted in \cite{Henkel:2002vd} that $alt_1$ arises mathematically as a contraction of (two copies of) the  Virasoro algebra though the contraction is slightly different from what we consider below.}}. So we hope that our explorations in the present paper would also be useful in the study of 
non-equilibrium statistical mechanical systems. 

It would be satisfying if one also had a Lagrangian description of systems with GCA. This has eluded us so far, and would continue to do so in our study of the two-dimensional case. But we hasten to remind the reader that Lagrangians, though useful, are not an essential characteristic for this symmetry to be realised in some context. It can very well be that the GCA has no Lagrangian realisation and in fact here we can draw analogies with the BMN limit where too the sector in $\mathcal{N}=4$ Supersymmetric Yang-Mills theory does not admit a lagrangian description.

A word about the original motivation of non-relativistic AdS/CFT and the case of GCA in two dimensions: recently, after the first version of this paper, in \cite{Hotta:2010qi}, it was found that the GCA in 2d could be realized as an asymptotic symmetry algebra of Cosmological Topologically Massive Gravity when the coefficient of the gravitational Chern-Simons term is scaled in a particular way. The central charges, that we would go on to describe in the 2D GCA have also been realized in the same scaling limit.

\subsection{Quantum GCA and the role of $D=2$}

The entire analysis in \cite{Bagchi:2009my} was classical and one would like to understand the quantum mechanical realisation of these symmetries better. This is specially true if one would like to exploit this symmetry in the context of a nonrelativistic limit of AdS/CFT as proposed in \cite{Bagchi:2009my}. A first step was taken in \cite{Bagchi:2009ca} (see also \cite{Alishahiha:2009np,
Martelli:2009uc}) where two and three point correlation functions (of primary fields) were obtained as solutions of the Ward identities for the finite part of the GCA (which arises as the contraction of $SO(d,2)$).\footnote{The expression for the two point function is actually contained in \cite{Henkel:2002vd} as part of more general expressions derived for arbitrary dynamical exponents.} These were found to be completely determined in form just as in the relativistic case. One would therefore imagine that the infinite dimensional GCA should give much stronger constraints when realised on the quantum dynamics just as in the case of relativistic $2d$ conformal invariance. 

With this in mind and given the centrality of $2d$ CFTs it is natural to specialise to two spacetime dimensions and study realisations of the infinite dimensional GCA in this case. In this process we might hope to extract dynamical information from the GCA comparable to that obtained from the Virasoro algebra. The main aim of this paper is to show that this hope can be largely realised. We find, in fact, that there is a very tight relation between the GCA symmetry in 2d and the relativistic Virasoro symmetry. Classically, we see that the infinite dimensional GCA in 2d arises as a contraction of the generators  of the usual holomorphic and antiholomorphic transformations. Quantum mechanically, in a 2d CFT, the latter get promoted to left and right moving Virasoro generators each with its central extension. We will find evidence that taking a scaling limit on a suitable family of 2d CFTs gives consistent quantum mechanical realisations of the GCA (now centrally extended), in what might be termed as 2d Galilean conformal field theories (or 2d GCFTs, for short). 

The families of 2d CFTs we will need to consider are rather unusual in that their left and right central charges $c$ and $\bar c$ are scaled (as we take the nonrelativistic limit) such that their magnitudes go to infinity but are opposite in sign. The parent theories are thus necessarily non-unitary and, not unsurprisingly, this non-unitariness is inherited by the daughter GCFTs. Since non-unitary 
2d CFTs arise in a number of contexts in statistical mechanics as well as string theory, one might expect that the 2d GCFTs realised here would also be interesting objects to study. 

\subsection{Strategy and outline of this paper}

Our study of 2d GCFTs in this paper proceeds along two parallel lines. The first line of development is as  described above and consists of taking carefully the non-relativistic scaling limit of the parent 2d CFT. We find that this limit, while unusual, appears to give sensible answers. Specifically, we will study in this way, the representation theory (including null vectors determined by a nonrelativistic Kac table), the Ward identities, fusion rules and finally the equations for the four point function following from the existence of level two null states. In all these cases we find that a non-trivial scaling limit of the 2d CFT's exists. This is not, {\it a priori}, obvious since, as we will see, the limit involves keeping terms both of 
${\cal O}({1\over \e})$ as well as of ${\cal O}(1)$(where $\e$ is the scaling parameter which is taken to zero).

The second line of development obtains many of these same results by carrying out an autonomous analysis of the GCA
i.e. independent of the above limiting procedure. In some cases we will see that the constraints from the GCA are slightly weaker than those arising from the scaling limit of parent relativistic CFTs. This is understandable since the limit of 2d CFTs is presumably only a particular way to realise GCFT's and there is no reason to expect it to be the only way to realise such GCFTs. Thus our dual considerations help to distinguish between some of the general results for any 2d GCFT from that obtained by taking the nonrelativistic limit.  It is also an important consistency check of our investigation that these two strands of development agree whenever they do and that it is always the GCA that gives weaker constraints.  

The structure of the paper is as follows: In the next section we discuss how, at the classical level, the generators of the $2d$ GCA arise from a group contraction of (combinations of) the usual holomorphic and anti-holomorphic vector fields. We also discuss the scaling limit one should take in 
the quantum theory thus relating the central charges of the GCA to the Virasoro central charges $c$ and $\bar c$. In Sec.~3 we proceed to construct representations of the $2d$ GCA in a manner analogous to the Virasoro representation theory, defining primaries and descendants. The primaries are labelled by a conformal weight $\Delta$ and a boost eigenvalue $\xi$. 
We also show that the state space is generically non-unitary. Sec.~4 deals with the non-relativistic Ward identities focussing on the case of two and three point functions. We briefly review the derivation of \cite{Bagchi:2009ca} and show how the answers there may also be obtained from the nonrelativistic scaling limit of the Virasoro algebra discussed in Sec.~2. 

We go back to the representation theory in Sec.~5 to consider null vectors 
of the GCA. We explicitly find the conditions for having null states at level two and check that the resulting conditions are precisely those obtained from the scaling limit of the Virasoro algebra. We go on to take the scaling limit of the Kac table for null states at arbitrary level and find a sensible nonrelativistic Kac table. From Sec.~6 onwards we focus on GCA primaries taking values in the nonrelativistic Kac table. We first derive the general differential equations for an $n$-point correlator which follow from the existence of level two null states and check that it is consistent with the form of the two point function of \cite{Bagchi:2009ca}. We proceed in Sec.~7 to derive the GCA fusion rules which follow from the GCA three point function using the differential equations of Sec.~6. These turn out to be only slightly weaker than what one would obtain from the scaling of the corresponding Virasoro fusion rules.

Finally, in Sec.~8 we consider the four point function 
which satisfies second order differential equations when one of the
primaries has a null descendant at level two. We can find the solution
in an explicit form both from this equation as well as from considering
the limit of the solution of the corresponding Virasoro four point
functions. The solution obeys various nontrivial conditions and seems to
be  consistent with the factorization into the three point function and the fusion rules. There is a  final section with remarks of a general nature. Appendix A discusses some issues regarding GCA descendants and their conformal blocks. Appendix B discusses a specific fusion channel in which the GCA gives a weaker constraint than the Virasoro. Appendix C contains various technical details relevant to the four point function in Sec.~8. 

\section{2d GCA from Group Contraction}

The maximal set of conformal isometries of Galilean spacetime generates the infinite dimensional Galilean Conformal
Algebra \cite{Bagchi:2009my}. The notion of Galilean spacetime is a little subtle since the spacetime metric degenerates 
into a spatial part and a temporal piece. Nevertheless there is a definite limiting sense (of the relativistic spacetime) in which one can define the conformal isometries (see \cite{Duval:2009vt}) of the nonrelativistic geometry. Algebraically, the set of vector fields generating these symmetries 
are given by
\ben{gcavec}
L^{(n)} &=& -(n+1)t^nx_i\p_i -t^{n+1}\p_t \,,\cr
M_i^{(n)} &=& t^{n+1}\p_i\,, \cr
J_a^{(n)} \equiv J_{ij}^{(n)} &= & -t^n(x_i\p_j-x_j\p_i)\,,
\een 
for integer values of $n$. Here $i=1\ldots (d-1)$ range over the spatial directions. 
These vector fields obey the algebra
\ben{vkmalg}
[L^{(m)}, L^{(n)}] &=& (m-n)L^{(m+n)}, \qquad [L^{(m)}, J_{a}^{(n)}] = -n J_{a}^{(m+n)}, \cr
[J_a^{(n)}, J_b^{(m)}]&=& f_{abc}J_c^{(n+m)}, \qquad  [L^{(m)}, M_i^{(n)}] =(m-n)M_i^{(m+n)}. 
\een
We expect that there would be a central extension to the Virasoro and Current algebras in the quantum theory. In fact, we will see that in two dimensions the GCA with central charges can be realised by taking a special limit of a relativistic 2d  CFT (albeit non-unitary). 

\subsection{GCA from Virasoro in 2d}

There is a  finite dimensional subalgebra  of the GCA (also sometimes
referred to as the GCA) which consists of taking $n=0,\pm1$ for the
$L^{(n)}, M_i^{(n)}$ together with $J_a^{(0)}$. This algebra is obtained
by considering the nonrelativistic  contraction of the usual (finite
dimensional) global conformal algebra $SO(d,2)$ (in $d>2$ spacetime
dimensions) 
(see for example \cite{negro1997}--\cite{Bagchi:2009my}). 

However, in two spacetime dimensions, as is well known, the situation is special. The relativistic conformal algebra is infinite dimensional and consists of two copies of the Virasoro algebra. One expects this to be also related, now to the infinite dimensional GCA algebra. Indeed in two dimensions the non-trivial generators in  \eq{vkmalg} are the $L_n$ and the $M_n$ (where we have dropped the spatial index from the latter since there is only one spatial direction and instead restored the mode number $n$ to the conventional subscript) :
\ben{gca2dvec}
L_n &=& -(n+1)t^n x\p_x -t^{n+1}\p_t\,, \cr
M_n &=& t^{n+1}\p_x\,,
\een
which obey
\ben{vkmalg2d}
[L_m, L_n] &=& (m-n)L_{m+n}\,, \qquad [M_{m}, M_{n}] =0\,, \cr
[L_{m}, M_{n}] &=& (m-n)M_{m+n}\,. 
\een

We will now show that the generators  in \eq{gca2dvec} arise precisely 
from a nonrelativistic contraction of the two copies of the Virasoro algebra of the relativistic theory.\footnote{This observation has also been independently made in \cite{IPM}. As mentioned in footnote 1, a slightly different contraction of the Virasoro algebra was also made in \cite{Henkel06} to obtain the same result.}
The non-relativistic contraction consists of taking the scaling 
\be{nrelscal}
t \rightarrow t\,, \qquad   x \rightarrow \epsilon x\,,
\ee
with $\epsilon \rightarrow 0$. This is equivalent to taking the velocities $v \sim \epsilon$ to zero
(in units where $c=1$).

Consider the vector fields which generate (two copies of) the centre-less Virasoro Algebra (or Witt algebra as it is often called) in two dimensions :
\be{repn2dV}
\L_n = -z^{n+1} \p_z\,, \quad \bL_n = -\z^{n+1} \p_{\z}\,.
\ee 
In terms of space and time coordinates, $z= t+x$, $\z=t-x$. Hence $\p_z = {1\over 2} (\p_t + \p_x)$ and $\p_{\z} = {1\over 2} (\p_t - \p_x)$. Expressing $\L_n , \bL_n$ in terms of $t,x$ and taking the above scaling  \eq{nrelscal} reveals that in the limit the combinations
 \ben{GCArepn}
\L_n + \bL_n &=& -t^{n+1}\p_t - (n+1)t^nx \p_x + \O(\e^2) \,,\cr
\L_n - \bL_n &=& -{1\over \e}t^{n+1} \p_x + \O(\e)\,.
\een 
Therefore we see that as $\e\rightarrow 0$
\be{Vir2GCA}
\L_n + \bL_n \longrightarrow L_{n}\,, \quad \e (\L_n - \bL_n) \longrightarrow - M_{n}\,.
\ee

Thus the GCA in 2d arises as the non-relativistic limit of the relativistic algebra. This was at the classical level of vector fields. At the quantum level the two copies of the Virasoro get respective central extensions
\ben{relalg}
[\L_m, \L_n] &=& (m-n) \L_{m+n} + {c \over 12} m(m^2-1) \delta_{m+n,0}\,, 
\nonumber \\[1mm]
[\bL_m, \bL_n] &=& (m-n) \bL_{m+n} 
+ {\bar c \over 12} m(m^2-1)\delta_{m+n,0}\,.
\een 
Considering the linear combinations \eq{GCArepn} which give rise to the GCA generators as in \eq{Vir2GCA}, we find 
\ben{gcawc}
[L_{m}, L_{n}] &=& (m-n) L_{m+n} + C_1 m(m^2-1) \delta_{m+n,0}\,, \crcr 
[L_{m}, M_{n}] &=& (m-n) M_{m+n} + C_2 m(m^2-1) \delta_{m+n,0}\,, \crcr 
[M_{m}, M_{n}] &=& 0\,.
\een
This is the centrally extended GCA in 2d.\footnote{One can check that a central extension in the commutator $[M_{m}, M_{n}]$ of the form $C_3(m)\delta_{m+n,0}$ is not allowed by the Jacobi 
identity.} 
Note that the relation between central charges is 
\be{centch}
C_1 = {{c+\bar c} \over 12}\,, \qquad {C_2 \over \e} = {{\bar c-c} \over 12}\,.
\ee
Thus, for a non-zero $C_2$
in the limit $\e\rightarrow 0$ we see that we need $\bar c-c \propto
\O({1\over \e})$. At the same time requiring $C_1$ to be finite we find
that  $c+\bar c$ should be $\O(1)$. 
(Motivated by the second equation
in (\ref{GCArepn}), we will make the slightly stronger assumption that
$\bar c - c = {\cal O}(1/\epsilon) + {\cal O}(\epsilon)$.)
Thus (\ref{centch}) can hold only if $c$ and $\bar c$ are large (in the limit $\e\rightarrow 0$) and opposite in sign. This immediately implies that the original 2d CFT on which we take the non-relativistic limit cannot be unitary. This is, of course, not a problem since there are many statistical mechanical models which are described at a fixed point by 
non-unitary CFTs.\footnote{We note that modular invariance of a 2d CFT implies that ${\bar c-c} \equiv 0  ({\rm mod} 24)$. We will implicitly assume that this is true for the CFT's we are considering. In other words, we will take ${C_2\over \e}$ to be an even integer. 
\label{ModularInv}
}

\section{Representations of the 2d GCA}

With these requirements in mind we now turn to the representations of the 2d GCA. We will be guided in this by the representation theory of the Virasoro algebra. 

\subsection{Primary States and  Descendants}

We will construct the representations 
by considering the states having definite scaling dimensions :
\begin{equation}
L_0 |\Delta \rangle = \Delta | \Delta \rangle \,.
\label{L0=Delta}
\end{equation}
Using the commutation relations (\ref{gcawc}), 
we obtain 
\begin{equation}
L_0 L_n | \Delta \rangle = (\Delta - n) L_n | \Delta \rangle, \quad 
L_0 M_n | \Delta \rangle = (\Delta - n) M_n | \Delta \rangle. 
\end{equation}
Then the $L_{n}, M_{n}$ with $n >0$ lower the value of the scaling
dimension, while those with $n<0$ raise it. If we demand that the
dimension of the states be bounded from below then we are led to
defining primary states in the theory having the following properties : 
\begin{equation}
L_n|\Delta \rangle_p =0\,, \quad 
M_n|\Delta \rangle_p =0\,,
\label{primop} 
\end{equation}
for all $n>0$. 
Since the conditions (\ref{primop})
are compatible with $M_0$ in the sense
\begin{equation}
L_n M_0 |\Delta \rangle_p = 0\,, \quad 
M_n M_0 |\Delta \rangle_p = 0\,,
\end{equation}
and also since $L_0$ and $M_0$ commute,
we may  introduce an additional label,
which we will call ``rapidity'' $\xi$:
\begin{equation}
M_0 |\Delta, \xi \rangle_p = \xi |\Delta, \xi \rangle_p\,.
\end{equation}

Starting with a primary state $|\Delta,\xi \rangle_p$, 
one can build up a tower of operators by the action of 
$L_{-n}$ and $M_{-n}$ with $n>0$. These will
be called the GCA descendants of the primary. 
The primary state together with its GCA descendants 
form a representation of GCA. 
As in the Virasoro case, we have to be 
careful about the presence of null states.
We will look at these in some detail
later in Sec.\ref{GCAnull}. 
An interesting property of the representation
constructed above is that a generic secondary 
state will not be an eigenstate of $M_0$. 
Examples of the secondary states, which are eigenstates
of $M_0$ are the ones which are constructed only with 
$M_{-n}$'s acting on the primary state.
In the rest of the paper, we omit the subscript ``$p$''
for the primary state.

The above construction is quite analogous to that of the 
relativistic 2d CFT. 
In fact, from the viewpoint of the limit (\ref{Vir2GCA})
we see that the two labels $\Delta$ and $\xi$ are related 
to the conformal weights in the 2d CFT as
\be{delxi}
\Delta=\lim_{\epsilon \to 0}
(h+\bar h)\,, \qquad  \xi= \lim_{\e\to 0} \e ({\bar{h} -h})\,,
\ee 
where $h$ and $\bar h$ are the eigenvalues of $\L_0$ and $\bL_0$, respectively. 
We will proceed to assume that such a scaling limit (as $\e \to 0$) of the 2d CFT 
exists. In particular, we will assume that the operator state correspondence 
in the 2d CFT gives a similar correspondence between the states and the operators
in the GCA :
\be{stateop}
\O (t,x) \leftrightarrow \O(0)|0\rangle\,, 
\ee
where $|0 \rangle$ would be the vacuum state
which is invariant under the generators 
$L_0, L_{\pm1}$, $M_0, M_{\pm1}$.  
Indeed in the rest of the paper, we will
offer several pieces of evidence that the scaling limit 
gives a consistent quantum mechanical system.

\subsection{Unitarity}

However already at level one of the GCA (using the usual terminology of CFT) 
we can see that the representation is generically non-unitary.
For this we have to define hermiticity and an inner product. 
We will assume that this is what would be naturally inherited 
from the parent relativistic CFT. In other words, we take
$L_n^{\dagger}=L_{-n}$ and 
$M_n^{\dagger}=M_{-n}$.\footnote{
Although the operator $M_0$ is Hermitian,
its action on the states is, in general,
not diagonalizable. 
This occurs for a special class of operators 
when the inner product is not positive definite.
See also the discussion around \eqref{nulleig}.
}

Let us now consider the tower above a primary state  
$|\Delta, \xi\rangle$. 
The states at level one are $L_{-1}|\Delta, \xi\rangle$ and
$M_{-1}|\Delta, \xi\rangle$. The (hermitian) matrix of 
inner products of these states can be evaluated and 
the determinant is easily seen to be given by $-\xi^2$. 
Thus, for non-zero $\xi$,  at least one of the eigenvalues 
must be negative. 
In such a case the Hilbert space clearly has negative norm
states. Thus only for $\xi=0$ might one have the possibility 
of a unitary representation though there are clearly null 
states even in this case.  
This is not too surprising given that the parent CFT 
is generically non-unitary as we argued above. 


\section{Non-Relativistic Ward Identities and Two and Three Point Functions} 

In \cite{Bagchi:2009ca} the GCA transformation laws 
for primary operators were written down as
(See eq. (4.3) and (4.4) of  \cite{Bagchi:2009ca}) :
\ben{primtrans}
\delta_{L_n} \O_p(t,x) &=& [L_n, \O_p(t,x)] \cr
&=& [t^{n+1}\p_t + (n+1)t^n x \p_x 
+ (n+1)(\Delta t^n - n \xi t^{n-1} x )] \O(t,x)\,, \cr
\delta_{M_n} \O_p(t,x) &=& [M_n, \O_p(t,x)] 
= 
[- t^{n+1}\p_x + (n+1) \xi t^n ] \O(t,x)\,.
\een
In \cite{Bagchi:2009ca} these transformation laws, in the special case of the global transformations (i.e. for $L_{0,\pm1}, M_{0,\pm1}$), were used to derive Ward Identities (see also \cite{Alishahiha:2009np, Martelli:2009uc}) for two and three point functions. The functional dependence of these correlators was then completely fixed by the resulting equations (see below for the explicit expressions). 
The analysis in \cite{Bagchi:2009ca} is based on the argument 
directly coming from GCA.
We will now see that all these statements can be also derived 
from the non-relativistic  limit of the relativistic 2d CFT.

\subsection{Transformation Laws}

Consider first how the transformation laws (\ref{primtrans}) 
of GCA primaries arise from the transformation laws of 
primary operators in 2d CFT, which are given by
\be{primV}
\tilde \O(\tilde z, \tilde \z) = \O (z, \z) {({dz \over d \tilde z})}^h {({d\z \over d \tilde \z})}^{\bar h}\,, 
\ee 
or infinitesimally under $z\rightarrow \tilde z =z+a(z)$ as 
\be{} 
\delta \O (z, \z) 
= - (h \p_z a(z) + a(z) \p_z) 
\O(z, \z )\,,
\ee
and similarly for $\bar z \to \tilde {\bar z} = \bar z + \bar a(\bar z)$ .
We take $a(z) = -z^{n+1}$ for $\delta_{{\cal L}_n}$
and $\bar a(\bar z) = - \bar z^{n+1}$ for $\delta_{\bar {\cal L}_n}$.

Motivated from the relation (\ref{Vir2GCA}), we may 
define the infinitesimal transformation $\delta_{L_n}$
and $\delta_{M_n}$ as
\begin{equation}
\delta_{L_n} {\cal O} 
= 
\lim_{\epsilon \to 0}
(\delta_{{\cal L}_n} + \delta_{\bar {\cal L}_n}) {\cal O}\,,
\qquad 
\delta_{M_n} {\cal O}
=
-
\lim_{\epsilon \to 0} \epsilon
(\delta_{{\cal L}_n} -  \delta_{\bar {\cal L}_n}) {\cal O}\,.
\end{equation}
Then by taking the limits 
on the right hand sides (RHS's), 
we obtain
\ben{ln}
\delta_{L_n} \O 
&=& \lim_{\e \to 0} [(n+1)h (t+ \epsilon x)^{n} 
+ 
{1\over 2} 
(t+ \epsilon x)^{n+1}
(\p_t + {1 \over \epsilon}\p_x) \nonumber\\ 
& & \quad + (n+1)\bar h (t-\epsilon x)^{n} + 
{1\over 2}(t- \epsilon x)^{n+1}(\p_t - {1 \over \epsilon }\p_x)] 
\O \nonumber\\
&=& [t^{n+1}\p_t + (n+1)t^n x \p_x 
+ (n+1)( \Delta t^n - n \xi t^{n-1} x )] \O\,,
\een
and similarly,
\ben{mn}
\delta_{M_n} \O 
&=& [- t^{n+1}\p_x + (n+1)t^n \xi] \O\,.
\een
where recall that $\Delta$ and $\xi$ are defined as (\ref{delxi}).
These are exactly those given in  \eq{primtrans}.

\subsection{Two and Three Point  Functions}

The constraints from the Ward identities for 
the global transformations $L_{0,\pm 1}, M_{0, \pm 1}$
apply to primary GCA operators.

Therefore consider the two point function of primary operators $\O_1(t_1, x_1)$ and $\O_2(t_2, x_2)$ of conformal and rapidity weights $(\Delta_1, \xi_1)$ and $(\Delta_2, \xi_2)$ respectively.
\be{2pt}
G_{\rm GCA}^{(2)}(t_1, x_1, t_2, x_2) 
= \langle \O_1(t_1,x_1) \O_2(t_2,x_2) \rangle \,.
\ee 
The correlation functions only depend on 
differences of the coordinates $t_{12} = t_1 - t_2$ 
and $x_{12} = x_1 - x_2$ because of the translation 
symmetries $L_{-1}$ and $M_{-1}$. 
The remaining symmetries give 
four more differential equations 
which constrain the answer to be 
\cite{Bagchi:2009ca}
\be{2ptgca}
G_{\rm GCA}^{(2)}(\{t_i, x_i \}) 
= C_{12} \delta_{\Delta_1,\Delta_2} 
\delta_{\xi_1, \xi_2} t_{12}^{-2\Delta_1} 
\exp\left( {2\xi_1 x_{12}\over t_{12}} \right).
\ee 
Here $C_{12}$ is an arbitrary constant,
which we can always take to be one
by choosing the normalization of the operators.

Similarly, the three point function of primary operators is given by 
\ben{3ptgca}
G_{\rm GCA}^{(3)}(\{t_i,x_i\}) 
& = &
C_{123}
t_{12}^{-(\D_{1}+\D_{2}-\D_{3})} 
t_{23}^{-(\D_{2}+\D_{3}-\D_{1})} 
t_{13}^{-(\D_{1}+\D_{3}-\D_{2})} \nonumber \\[1mm]
&& \hspace{-2cm}\times 
\exp\left( 
{\frac{(\xi_1+ \xi_2 -\xi_3)x_{12}}{t_{12}}}
+ 
{\frac{(\xi_2+ \xi_3 -\xi_1)x_{23}}{t_{23}}}  
+ 
{\frac{(\xi_1+ \xi_3 -\xi_2)x_{13}}{t_{13}}}
\right) \,, 
\een
where $C_{123}$ is an arbitrary constant. 
So we see that like in the case of relativistic CFTs, 
the three point function is fixed upto a constant.\footnote{
See, however, the explanation in the footnote \ref{relativesign}.}

\subsection{GCA Correlation Functions from 2d CFT}
\label{GCA CF}

We now show that these expressions for the GCA two and three point 
functions can also be obtained by taking an appropriate scaling 
limit of the usual 2d CFT answers.
This limit requires scaling the quantum numbers
of the operators as (\ref{delxi}), 
along with the non-relativistic limit
for the coordinates \eq{nrelscal}.

Let us first study the scaling limit of the 
two point correlator.\footnote{This was obtained in discussion with S. Minwalla.}
\ben{R2pt}
G^{(2)}_{\rm 2d\,CFT} 
&=& \delta_{h_1, h_2} \delta_{{\bar h}_1, {\bar h}_2}  
z_{12}^{-2 h_1} \bar z_{12}^{-2 \bar h_1} 
\nonumber \\
&=& 
\delta_{h_1,h_2}\delta_{\bar h_1,\bar h_2}\,\,
t_{12}^{-2h_1} 
\Big( 1 + \epsilon {x_{12} \over t_{12}} \Big)^{-2h_1} 
\,t_{12}^{-2\bar h_1} 
\Big(1 - \epsilon {x_{12} \over t_{12}} \Big)^{-2 {\bar h}_1}
\nonumber \\
&=& \delta_{h_1, h_2} \delta_{{\bar h}_1, {\bar h}_2} \,\ 
t_{12}^{- 2(h_1+\bar h_1)} 
\exp\Big( 
- 2 (h_1- \bar h_1) 
\big(\epsilon { x_{12} \over t_{12}} 
+ {\cal O}(\epsilon^2)\big)\Big)\,.
\een
Now by taking the scaling limit as (\ref{delxi}), 
we obtain the GCA two point function
\be{lim2pt} 
\lim_{\epsilon \to 0} G^{(2)}_{\rm 2d\,CFT} 
=  \delta_{\Delta_1, \Delta_2} 
\delta_{{\xi}_1, {\xi}_2} \, t_{12}^{-2\Delta_1} 
\exp \Big({2 \xi_1 x_{12} \over {t_{12}}} \Big) 
= G^{(2)}_{\rm GCA}\,.
\ee

A similar analysis yields the three point 
function of the GCA from the relativistic 
three point function.
The relativistic three point function is written as
\begin{eqnarray}
G^{(3)}_{\rm 2d\,CFT} 
& = &
C_{123} 
z_{12}^{-(h_1 + h_2 - h_3)} 
z_{23}^{-(h_2 + h_3 - h_1)} 
z_{13}^{-(h_1 + h_3 - h_2)} 
\times\,\, (\textrm{anti-holomorphic}) 
\label{rel3pt-1}\\[2mm]
& = & C_{123}
t_{12}^{-(h_1 +  h_2  - h_3)} 
t_{12}^{-(\bar h_1 +  \bar h_2  - \bar h_3)} 
{\rm e}^{-  (h_1 + h_2 - h_3)
(\epsilon {x_{12}\over t_{12}} 
+ {\cal O}(\epsilon^2))}
{\rm e}^{(\bar h_1 + \bar h_2 - \bar h_3)
(\epsilon {x_{12}\over t_{12}}+  {\cal O}(\epsilon^2))} \nonumber  \\[1mm]
& &\hspace{5cm} \times \,\,(\textrm{product of two permutations})\,.
\label{rel3pt}
\end{eqnarray}
Then taking the non-relativistic limit,
we obtain the GCA three point function \eq{3ptgca}.
Note that the constant factor in (\ref{3ptgca}) would
be given by taking the limit of the constant in (\ref{rel3pt-1}).

We should also mention here the issue of singlevaluedness of the correlation functions.
The holomorphic part, or the anti-holomorphic part, of the 
correlation functions in the 2d CFT have branch cuts 
for generic values of 
the conformal dimensions $h_i$ and $\bar h_i$.
Then the requirement for the singlevaluedness 
of the two point functions gives rise to 
a condition that the difference $h_i - \bar h_i$ 
of every operator must be an integer or a half integer.
This is the usual spin statistics theorem.\footnote{
A word regarding  the overall phase of correlation functions:
If we pay attention to the overall phase 
in the derivation \eq{R2pt}, it would be like ($x=0$ for simplicity)
$z^{-2 h} \bar z^{-2 \bar h} 
\to (\pm1)^{2(h - \bar h)} |t|^{-2(h+\bar h)}$.
Here the plus (the minus) sign is for $t>0$ ($t<0$),
and the exponent $2(h - \bar h)$ is the twice of the spin as mentioned 
in the main text. Note that in the derivation of 
\eq{2ptgca}, the differential equation is solved 
separately for the two segments $t < 0$ and $t > 0$.
So far we have no argument, from the GCA side,
to fix the relative 
coefficient of the solutions for these segments.
However, when we take the non-relativistic limit from 2d CFT,
this ambiguity is fixed as we saw above. 
This relative phase plays an important role in the 
discussion of the four point function.
\label{relativesign}
} 
We see that this theorem need not hold in a generic GCFT
since there is no such singlevaluedness requirement. 
However, GCFT's arising as limits of 2d CFTs would inherit 
this relation.


\section{GCA Null Vectors}
\label{GCAnull}
Just as in the representation of the Virasoro algebra, 
we will find that there are null states in the GCA tower built on a
primary  $|\Delta, \xi\rangle$ for special values of $(\Delta,\xi)$.  
These are states which are orthogonal to all states in the tower
including itself.  
We can find the null states at a given level by writing the most general
state at that level as a combination of the  $L_{-n}, M_{-n}$'s $(n>0)$
acting on the GCA primary and then imposing the condition that all the
positive modes $L_n, M_n$ (with $n>0$) annihilate this state. Actually
one needs  to only impose this condition for $n=1$ and $n=2$
since the others are given as the commutators of these modes.
This will give conditions that fix the relative coefficients in the
linear combination as well as give a relation between  
$\Delta,\xi$ and the central charges $C_1, C_2$. 

Thus at level one we have only the states $L_{-1}|\Delta, \xi\rangle$ and $M_{-1}|\Delta, \xi\rangle$. 
It is easy to check that one has a null state only if $\xi$ is zero. 
At level two things are a little more non-trivial. Let us consider the most general level two state of the form 
\be{levtwo}
|\chi \rangle = (a_1L_{-2}+a_2L_{-1}^2+b_1L_{-1}M_{-1}+d_1M_{-1}^2+d_2M_{-2})|\Delta, \xi\rangle .
\ee
We now impose the condition that $L_{1,2}, M_{1,2}$ annihilate this state. 

A little algebra using \eq{gcawc} gives us the conditions :
\begin{align}
&3a_1+2(2\Delta+1)a_2+ 2\xi b_1 = 0; 
\quad (4\Delta+ 6C_1)a_1+6 \Delta a_2 + 6\xi b_1+(6C_2+4\xi)d_2 =0; 
\cr
& 2(\Delta+1)b_1+4 \xi d_1+3d_2 = 0; 
\quad (4\xi+6C_2)a_1 =0; 
\quad 3a_1+2a_2+2 \xi b_1 =0; \quad \xi a_2  =0. 
\label{levtwocond}
\end{align}
We will now separately consider the two 
cases where $C_2\neq 0$ and $C_2=0$. 

\subsection{The Case of $C_2\neq 0$}

Here we will first consider the case where $\xi\neq 0$. In this case we have two further 
options. Either  $a_1=0$ or $a_1\neq 0$. In the former case, $b_1=0$ as well. We then find that for a nontrivial solution,   
$\xi=-{3C_2\over 2}$ and  $d_1=-{3\over 4\xi}d_2$. Thus there is a null state of the form
\be{null1}
|\chi^{(1)} \rangle =(M_{-2}-{3\over 4\xi}M_{-1}^2)|\Delta, \xi\rangle .
\ee

We can also consider solutions for which $a_1\neq 0$. In this case $b_1= -{3\over 2\xi}a_1$. We
have once again $\xi=-{3C_2\over 2}$.
But then we also have to satisfy the consistency condition $\Delta={(9-6C_1)\over 4}$. 
For $\Delta\neq -1$, we also must have at least one of $d_1,d_2$ nonzero for such a solution. By taking a suitable linear combination with $|\chi^{(1)} \rangle$ we can choose $d_2=0$
 and then we get another null state of the form
\be{null2}
|\chi^{(2)} \rangle =(L_{-2}-{3\over 2\xi}L_{-1}M_{-1} +{3(\Delta+1)\over 4\xi^2}M_{-1}^2)|\Delta, \xi\rangle.
\ee

We can also have the case where $\xi=0$. Then we must have $a_1=a_2=d_2=0$ and $d_1$ is again undetermined corresponding again to $|\chi^{(1)} \rangle = M_{-1}^2|\Delta, 0\rangle$. However, it is only for $\Delta=-1$ that one gets a second null state. Here $b_1$ is undetermined and corresponds to 
$|\chi^{(2)} \rangle =  L_{-1}M_{-1}|\Delta=-1, 0 \rangle$. Note, however, that both these states are descendants of a level {\it one} null state $M_{-1} |\Delta, 0\rangle$.

In any case, for both $\xi=0$ as well as $\xi\neq 0$, the null states obey
\be{nulleig}
M_{0}|\chi^{(2)} \rangle =\xi |\chi^{(2)} \rangle +\alpha |\chi^{(1)} \rangle, \qquad 
M_{0}|\chi^{(1)} \rangle = \xi|\chi^{(1)} \rangle,
\ee
where $\alpha$ is either of $1,2$.
That is, they are eigenstates of $M_0$ (upto null states). 
More generally, this is a reflection of the fact that the operator $M_0$ is not diagonal on the states 
at a given level. In fact, it can be easily seen to take an upper triangular form in terms of a suitable ordering of the basis elements at a given level. This is reminiscent of the behaviour in logarithmic CFTs. 

At a general level $K$ one will find in parallel with $|\chi^{(1)} \rangle$ a null state  formed only from the $M_{-n}$'s taking the form
\be{genull}
|\chi \rangle = \left( M_{-K} + \eta_1 M_{-K+1} M_{-1} + \ldots \right)  |\Delta, \xi \rangle.
\ee
The action of $L_K$ gives the constraint as
\begin{equation}
L_K |\chi \rangle 
= \left( [L_K, M_{-K}] + \ldots \right)  |\Delta, \xi \rangle 
\quad \to \quad 2K \xi + C_2 K(K^2-1) =0.
\label{nullkcond}
\end{equation}
Here we have used the fact that all other commutators generate $M_m$ with $m>0$ and these go through and annihilate the state $ |\Delta, \xi \rangle$.  Thus there are null states at level $K$ if 
$\xi$ obeys the above relation with respect to $C_2$. 

\subsection{The Case of $C_2=0$}

In this case, it is easy to see that we have a non-zero null state only if $\xi=0$. Then from \eq{levtwocond} we find that for $(\Delta \neq -{3\over 2}C_1)$ that both $a_2$ and  
$a_1=0$ and we have a relation $d_2=-{2(\Delta+1)\over 3}b_1$ and $d_1$ is undetermined. Therefore the two independent null states 
are now 
\be{nullxi0}
|\chi^{(1)} \rangle = M_{-1}^2|\Delta, 0\rangle,
\ee{}
(which is just a descendant of the level one null state) and 
\be{nullxi}
|\chi^{(2)} \rangle =  (L_{-1}M_{-1}-{2(\Delta+1)\over 3}M_{-2})|\Delta, 0 \rangle.
\ee
Note that there is no constraint on the  values of  $C_1, \Delta$ apart from the fact that $(\Delta \neq -{3\over 2}C_1)$. In fact, in the case of $(\Delta = -{3\over 2}C_1)$ we find that we can have $a_1,a_2\neq 0$ only if $\Delta=C_1=0$, which is a trivial case. So we continue to take $a_1=a_2=0$  even when $(\Delta = -{3\over 2}C_1)$ and we have the two null states  $|\chi^{(1,2)} \rangle$ for all values of $\Delta, C_1$.

We can actually say more about these states. Note that the states
$M_{-1}^2|\Delta, 0\rangle$ and $L_{-1}M_{-1}|\Delta, 0\rangle$ are descendants of the level one 
null state $M_{-1}|\Delta, \xi\rangle$. 
Thus if we consistently set the null state at level one to zero together with its descendants, then the new null state at level two is given by $M_{-2}|\Delta, 0\rangle$. 
Continuing this way it is easy to see that we have a new null state given by $M_{-K}|\Delta, 0\rangle$ 
at level $K$ if we set all the null states (and their descendants) at lower levels to zero. 
In this case, the GCA tower precisely reduces to the Virasoro tower given by the Virasoro 
descendants of the primary. As noted for level two, there is generically no condition on $C_1, \Delta$ : we only require $\xi=C_2=0$. Thus we can consider a truncation of the Hilbert space to this Virasoro module. 
We can, of course, have the Virasoro tower reducible by having Virasoro null vectors. These are analysed in the usual way. Note that we can have unitary representations if $\Delta$ and $C_1$ obey the conditions familiar from the study of the Virasoro algebra. 

Unfortunately, this sector is relatively uninteresting from the point of view of its spacetime dependence. The correlation functions of operators are ultralocal depending only on time. This is because all $x$ dependence arises in combination with the $\xi$ dependence so as to survive the nonrelativistic limit. 
Setting $\xi=0$ thus removes the spatial dependence of correlators.  
  
\subsection{GCA Null Vectors from 2d CFT}

Here we will show that the level two GCA null vectors can alternatively be obtained 
by taking the nonrelativistic scaling limit of the familiar level two null vectors of 2d CFT. 

The null vector at level two in a Virasoro tower is given by
\be{relnull}
|\chi_L \rangle = ({\cal L}_{-2} + \eta {\cal L}_{-1}^2) 
|h \rangle \otimes |\bar h \rangle \,,
\ee
with
\be{nullcond}
\eta =  - {3 \over 2(2h+1)},
\ee
\be{nullcond2}
h = {1 \over 16}
( 5 - c \pm \sqrt{(1-c)(25-c)}).
\ee
One has a similar null state for the antiholomorphic Virasoro
obtained by replacing
${\cal L}_n \to \bar {\cal L}_n$, $h \to \bar h$
and $c \to \bar c$.

Using these expressions let us first take the limit of the relation in 
\eq{nullcond2} (together with its antiholomorphic counterpart). Recall that the nonrelativistic scaling limit for the central charges and conformal weights are given by \eq{centch} and \eq{delxi}.
Since the central charges $c,\bar{c}$ are opposite in sign as
$\epsilon \to 0$, on taking the positive sign for the negative
central charge part,\footnote{If we take the negative sign in the square
root we get in the same limit where $C_2\neq 0$ that $\xi=0$ and
$\Delta=-1$. This is precisely what we obtained in Sec.~5.1.
\label{xi=0Delta=-1}} and vice
versa, for the square root in \eq{nullcond2}, we get 
\be{nreldelxi}
\xi = 
-{3C_2\over 2}\,, \qquad
\Delta = {(9-6C_1)\over 4}\,.
\ee
These are precisely the relations we obtained in the previous section if we require the existence of both the GCA null states $|\chi^{(1)} \rangle, |\chi^{(2)} \rangle$ at level two.

These states themselves can be obtained by taking the nonrelativistic limit on appropriate combinations of the relativistic null vectors $|\chi_L \rangle$ and its antiholomorphic counterpart $|\chi_R\rangle$. 
Consider
\be{nullplus}
|\chi^{(1)} \rangle = \lim_{\e\to 0} {\epsilon}(- |\chi_L \rangle  + |\chi_R \rangle) \,, \qquad |\chi^{(2)} \rangle = \lim_{\e \to 0}  (|\chi_L \rangle  + |\chi_R \rangle) .
\ee
From the expressions \eq{delxi}, we obtain $\eta={3\epsilon\over 2\xi}(1+{(\Delta+1)\epsilon\over \xi})$
and $\bar{\eta}= -{3\epsilon\over 2\xi}(1-{(\Delta+1)\epsilon\over \xi})$ upto terms of order $\epsilon^2$. 
Substituting this into \eq{nullplus}, using the relations \eq{Vir2GCA} and taking the limit $\e\to 0$, we obtain
\ben{nulllim}
|\chi^{(1)} \rangle &=& (M_{-2}-{3\over 4\xi}M_{-1}^2)|\Delta, \xi\rangle, \cr
|\chi^{(2)} \rangle & =& (L_{-2}-{3\over 2\xi}L_{-1}M_{-1} +{3(\Delta+1)\over 4\xi^2}M_{-1}^2)|\Delta, \xi\rangle,
\een
which are exactly what we found from the intrinsic GCA analysis  in \eq{null1} and \eq{null2}.

Similarly the case mentioned in footnote \ref{xi=0Delta=-1} is also easily seen to correspond to the null states constructed in Sec.~5.1 for $C_2\neq 0$. Finally, there is the case when $C_2=0$ and hence $c=\bar c$. Therefore it follows from \eq{nullcond2} and its antiholomorphic counterpart that $h=\bar h$, i.e., $\xi=0$ and $\Delta=2h$. It is then easy to verify that the pair of states in \eq{relnull} and its antiholomorphic counterpart reduce in the non-relativistic limit to the states constructed in Sec.~5.2 (see \eq{nullxi0} and \eq{nullxi}). 
It is satisfying that the limiting process gives answers consistent with the intrinsic GCA analysis. 

\subsection{Non-Relativistic Limit of the Kac Formula}

More generally if we want to examine the GCA null states at a general level, we would have to 
perform an analysis similar to that in the Virasoro representation theory. A cornerstone of this analysis is the Kac determinant which gives the values of the weights of the Virasoro Primaries $h (\bar h)$ for which the matrix of inner products at a given level has a zero eigenvalue :
\be{kd}
\mbox{det} M^{(l)} = \alpha_l \prod_{1\leq r,s ; rs \le l} (h - h_{rs}(c))^{p(l-rs)}\,,
\ee
where $\alpha_l$ is a constant independent of $(h,c)$; $p(l-rs)$ is the number of partitions of the integer $l-rs$. 
The functions $h_{r,s}(c)$ are expressed in a variety of ways. One convenient representation is :
\ben{h-rep1}
h_{r,s}(c) &=& h_0 + {1 \over 4}(r \alpha_+ + s \alpha_-)^2\,, \\
h_0 &=& {1 \over {24}}(c-1)\,, \\
\alpha_{\pm} &=& {\sqrt{1-c} \pm \sqrt{25 -c} \over {\sqrt{24}}}\,.
\een
One can write a similar expression for the antiholomorphic sector. 
The values $h_{r,s}$ are the ones for which we have zeroes of the determinant and hence null vectors (and their descendants). 

One could presumably generalise our analysis for GCA null vectors at level two and directly obtain the GCA determinant at a general level. This would give us a relation for $\Delta$ and $\xi$ in terms of 
$C_1, C_2$ for which there are null states, generalising the result
\be{nutwores}
\xi=-{3C_2\over 2}\,, \qquad \Delta ={(9-6C_1)\over 4}\,,
\ee
at level two.
However, here instead of a direct analysis we will simply take the non-relativistic limit of the Kac formula
and see that one obtains sensible expressions for the  $\Delta$ and $\xi$ at which the GCA determinant would vanish. 

In taking the non-relativistic limit, let us first consider the case where $C_2 \neq 0$ and chosen to be positive. Therefore (from \eq{centch}) we need to take 
$c \ll -1$ and $\bar c \gg 1$ as $\e \to 0$. We then find
\begin{eqnarray}
h_{r,s} &=& {1 \over 24} c(1-r^2) + {1 \over 24} (13 r^2 -12 rs -1) 
+ {\cal O}(\epsilon), 
\label{hh1} \\
\bar h_{r',s'} &=& {1 \over 24} \bar c (1-r^{\prime 2}) 
+ {1 \over 24} (13 {r'}^2 -12 r's' -1) + {\cal O}(\epsilon).
\label{hh2}
\end{eqnarray}
Then eq.  \eq{delxi} gives the values of 
$\Delta$ and $\xi$ in terms of the RHS of 
\eq{hh1} and \eq{hh2} which in turn can be 
expressed in terms of $C_1$ and $C_2$.
In the simple case where we take $(r,s)=(r',s')$\footnote{
Requiring that $\Delta$ should not have a 
$\frac{1}{\epsilon}$ piece immediately implies that $r=r'$.
The choice $s=s'$ is merely to simplify expressions.
\label{r=r's=s'}}
\begin{eqnarray}
\Delta_{r,s} &=& 
\lim_{\epsilon \to 0} (h_{r,s} + \bar h_{r,s}) 
= {1 \over 2}C_1 (1-r^2) + {1 \over 12} (13 r^2 -12 rs -1) \,,
\label{null}\\
\xi_{r,s} &=& - \lim_{\e\to 0} {\e}(h_{r,s} - \bar h_{r,s}) 
= {1 \over 2} C_2(1-r^2)\,.
\label{xirs}
\end{eqnarray}

In the case of $r=2$, $s=1$ we have two null states 
at level two built on the primary 
$|\Delta_{2,1}, \xi_{2,1} \rangle$. 
We see from \eq{null} and \eq{xirs} that 
the values of $\Delta_{2,1}$ and $\xi_{2,1}$ 
are exactly those given in \eq{nutwores}. 
This is also what we explicitly constructed in the previous subsection.

We can also construct a pair of null states at level two for $r=1, s=2$. In this case, we see that $\Delta_{1,2}=-1$ and $\xi_{1,2}=0$. Such a null state for $C_2\neq 0$ was found in Sec. 5.1 and we see that it corresponds to the case mentioned in footnote \ref{xi=0Delta=-1}. 

Finally, as a last consistency check, we note that at level $K$, we can
have a null state with $r=K, s=1$. The condition on $\xi_{K,1}$ given by
\eq{xirs} is exactly the same as given in  
\eq{nullkcond}.

\section{Differential Equations for GCA Correlators from Null States}
\label{DiffEqGCA}
The presence of the null states gives additional relations between correlation functions which is at the heart of the solvability of relativistic (rational) conformal field theories. To obtain these relations one starts with differential operator realisations $\hat \L_{-k}$ of the $\L_{-k}$ with $(k \geq 1)$. 
Thus one has
\be{hatl}
\langle (\L_{-k}\phi (z,\bar z)) 
\phi_1(z_1,\bar z_1) \cdots \phi_p(z_n,\bar z_n) \rangle
= \hat \L_{-k} \langle \phi (z,\bar z)
\phi_1(z_1,\bar z_1) \cdots \phi_p(z_n,\bar z_n) \rangle \,,
\ee
where
\begin{equation}
\hat \L_{-1} = \partial_z\,, \quad 
\hat \L_{-k} = \sum_{i=1}^n
\left\{
{(k-1) h_i\over (z_i - z)^k}
-
{1 \over (z_i - z)^{k-1}} \partial_{z_i}
\right\} \quad (\mbox{for} \,k \geq 2).
\label{calL(-k)}
\end{equation}

We can obtain similar differential operators for the GCA generators by taking appropriate 
non-relativistic limits \eq{nrelscal} of these  relativistic expressions. Expanding the operators
$\hat \L_{-k}$ as 
\begin{equation}
\hat \L_{-k} 
= 
\epsilon^{-1} \hat \L_{-k}^{(-1)}
+
\hat \L_{-k}^{(0)}
+
{\cal O}(\epsilon)\,,
\end{equation}
and similarly for the anti-holomorphic part.
Using \eq{Vir2GCA}, we obtain the expressions for the differential operators $\hat M_{-k}$ and $\hat L_{-k}$ as :
\ben{Mm}
\hat M_{-k} & =& \sum_{i=1}^{n} [\frac{(k-1)\xi_i}{t^{k}_{i0}} + \frac{1}{t^{k-1}_{i0}} \partial_{x_i}] \,, \nonumber \\
\hat L_{-k} & = & \sum_{i=1}^{n} [\frac{(k-1)\Delta_i}{t^{k}_{i0}} + \frac{k(k-1)\xi_i}{t^{k+1}_{i0}} x_{i0} - \frac{1}{t^{k-1}_{i0}}\partial_{t_i} + \frac{k-1}{t^{k}_{i0}} x_{i0} \partial_{x_i} ] \,, 
\een{}
where $x_{i0} = x_i - x$ and $t_{i0} = t_i - t$. 
For $k=1$ we have the simpler expressions $\hat M_{-1}=-\partial_x$ and $\hat L_{-1}=\partial_t$.

Therefore, correlation functions of GCA descendants of a primary field
are given in terms of the correlators of the primaries by the action of
the corresponding differential operators $\hat M_{-k}$ and $\hat
L_{-k}$. 
One can derive the same result from intrinsic GCA
analysis by following the method used to derive eq~(13) in \cite{Qiu}.
In appendix A we illustrate how this works and give the analogue of the
conformal blocks for the non-relativistic case.

Now we will study the consequences of having null states at level two. We will consider the two null states  $|\chi^{(1)} \rangle \,, \,|\chi^{(2)} \rangle$ of Sec. 5.1, or rather correlators involving the corresponding fields $\chi^{(1,2)}(t,x)$. Setting the null state and thus its correlators to zero
gives rise to differential equations for the correlators involving the
primary $\phi_{\Delta,\xi}(t,x)$ with other fields. Using the forms
 \eq{null1} and \eq{null2}, we find that the differential equations take the
 form 
\begin{align}
&(\hat M_{-2}-{3\over 4\xi}\hat M_{-1}^2) 
\langle \phi_{\Delta,\xi} (t,x)
\phi_1(t_1, x_1) \cdots \phi_n(t_n, x_n) \rangle =0\,, 
\label{nulleq1} \\
&(\hat L_{-2}-{3\over 2\xi}\hat L_{-1} \hat M_{-1} 
+{3(\Delta+1)\over 4\xi^2}\hat M_{-1}^2) \langle 
\phi_{\Delta,\xi} (t,x)
\phi_1(t_1,x_1) \cdots \phi_n(t_n,x_n) \rangle =0\,,
\label{nulleq2}
\end{align}
with $\hat L_{-2}$ and $\hat M_{-2}$ as given in equation  \eq{Mm}. 

Acting on a two point function 
\be{twopt}
G^{(2)}_{\rm GCA}(t,x)=\langle \phi_{\Delta,\xi} (t,x) \phi_{\Delta^{\prime},\xi^{\prime}}(0, 0)  \rangle \,,
\ee
we have the simple differential equations
\ben{twoptnull}
&& [{\xi^{\prime} \over t^2}+{1\over t}\partial_x-{3\over
4\xi}\partial_x^2]G_{\rm GCA}^{(2)}(t,x) = 0\,,\\[1mm]
&& 
[ {\Delta^{\prime} \over t^2}
+2{\xi^{\prime}x\over t^3}
-{1\over t}\partial_t
+{x\over t^2}\partial_x
+{3\over 2\xi}\partial_t\partial_x+{3(\Delta+1)\over 4\xi^2}\partial_x^2
]G^{(2)}_{\rm GCA}(t,x) = 0 \,.
\een

It is not difficult to check that the GCA two point function \cite{Bagchi:2009ca} given in \eq{2ptgca} 
$G^{(2)}_{\rm GCA}(t,x) \propto t^{-2 \Delta} {\rm e}^{2\xi x \over t}$ 
(together with $\xi=\xi^{\prime}$ and $\Delta=\Delta^{\prime}$) identically 
satisfies, as it should, both these differential equations.

\section{GCA Fusion Rules}

Analogous to the relativistic case
\cite{Ketov:1995yd}, 
we can derive "\textit{Fusion rules}",
\be{fus}
[\phi_1] \times [\phi_2] \simeq \sum_{p} [\phi_p] \,\, ,\nonumber 
\ee
for the GCA conformal families, that determine which families $[\phi_p]$ have their primaries and descendants occurring in an OPE of any two members of the families $[\phi_1]$ and $[\phi_2]$. Here we have denoted a family $[\phi_i]$ by the corresponding primary $\phi_i$.

We illustrate how the fusion rules can be obtained 
for the families $[\phi_{\Delta , \xi}]$ and $[\phi_{\Delta_1 , \xi_1}]$, where both fields are members of the GCA Kac table as 
specified by \eq{hh1} \eq{hh2}. 
As mentioned in footnote \ref{r=r's=s'}, we need to take $r=r'$.
The resulting $\Delta,\xi$ are thus labelled by 
a triple $\lbrace r (s,s') \rbrace$. 
In particular, we will consider below the case of
 $\Delta =\Delta_{2 (1,1)}$ and $\xi=\xi_{2(1,1)}$.
(In Appendix B, we consider the case of the fusion rule following from
the case where 
$\Delta =\Delta_{1 (2,2)}$ and $\xi=\xi_{1(2,2)}$. This case is
interesting in that the GCA limit  
generally gives a weaker constraint than that following from the nonrelativistic limit of the 2d CFT.)

The fusion rules are derived from applying the condition that
$\phi_{\Delta , \xi}$ has a null descendant at level two. For ($\Delta_1,\xi_1$) we will consider a general member $r (s,s')$ of
the GCA Kac table.  
Thus we have from \eq{nreldelxi}, \eq{hh1} and \eq{hh2} :
\begin{align}
\Delta &= \Delta_{2(1,1)} 
=  \frac{1}{4} (9 - 6C_1)\,, \quad 
\xi =\xi_{2(1,1)} 
= -\frac{3C_2}{2} \,;
\label{fus1} \\
\Delta_1 &= \Delta_{r(s,s^{\prime})}
= {C_1\over 2}(1-r^2)+ \frac{1}{12} 
\lbrace 13 r^2  - 6 r (s + s^{\prime})-1\rbrace \,, 
\label{fus1-2} \\ 
\xi_1 &= \xi_{r(s, s^{\prime})}= \frac{C_2}{2} (1 - r^2) \,.
\label{fus1-3}
\end{align}

We need to consider the conditions \eq{nulleq1} and \eq{nulleq2} for the
case of the three point function, i.e., $n=2$. 
With $G^{(3)}_{\rm GCA}
(t,x,\{t_i,x_i\})=\langle \phi_{\Delta,\xi}(t,x) 
\phi_{\Delta_1,\xi_1}(t_1,x_1)
\phi_{\Delta_2,\xi_2}(t_2,x_2)\rangle$,
these give the constraints :
\begin{eqnarray}
&&[ \sum_{i=1}^{2} (\frac{\xi_i}{t^{2}_{i0}} + \frac{1}{t_{i0}}
\partial_{x_i} ) - \frac{3}{4 \xi} \partial_{x}^2 ] \,\, G^{(3)}_{\rm GCA}
= 0 \label{nullcons1}\,, \\
&&[\sum_{i=1}^{2} ( \frac{\Delta_i}{t^{2}_{i0}} + \frac{2\xi_i}{t^{3}_{i0}} x_{i0} - \frac{1}{t_{i0}}\partial_{t_i} + \frac{1}{t^{2}_{i0}} x_{i0}\partial_{x_i})
 + \frac{3}{2 \xi} \partial_{x} \partial_t + \frac{3}{4} \frac{\Delta
 + 1}{\xi^2} \partial_{x}^2 ] G^{(3)}_{\rm GCA} \,\,
= 0 \,,
\label{nullcons2}
\end{eqnarray}
respectively.
Now by using (\ref{3ptgca}), these translate into
\ben{hogehoge} 
&&{1\over 2}\xi + (\xi_1+\xi_2) - \frac{3}{2 \xi}(\xi_1- \xi_2)^2 =0\,, \cr
&& 4(2\Delta_1 -  \Delta_2 +\Delta) + \frac{3 (\xi_2 - \xi
-\xi_1)^2}{\xi^2} (\Delta +1)  - \frac{6(\xi_2 - \xi
-\xi_1)}{\xi}(\Delta_2-\Delta_1-\Delta - 1) = 0\,. \nonumber 
\een

Solving the above equations, we get two simple sets of solutions :
\be{soln1}
\xi_2 = \frac{C_2}{2} [1 - (r\pm 1)^2] \,, \quad \Delta_2 =  {1\over 2}C_1 \lbrace 1 - (r \pm 1)^2\rbrace+ \frac{1}{12} [ 13(r\pm 1)^2  - 6 (r\pm1) (s + s^{\prime})-1 ].
\ee{}
Comparing with \eq{fus1-2} and \eq{fus1-3}, we see that 
\be{deltwofus}
\Delta_2=\Delta_{r\pm1(s,s^{\prime})}\,, \qquad \xi_2=\xi_{r\pm1(s,s^{\prime})}\,,
\ee
which is exactly what the relativistic fusion rules imply, namely
\be{spfus}
[\phi_{2(1,1)}] \times [\phi_{r(s,s^{\prime})}] =  [\phi_{r+1(s,s^{\prime})}] +[\phi_{r -1(s,s^{\prime})}]\,.
\ee
Thus once again we see evidence for the consistency of the GCA limit of
the 2d CFT. In this case the GCA analysis gives as strong a constraint
as the relativistic CFT. 
However, as mentioned above, in Appendix B we will 
give an example where the GCA analysis gives a weaker
constraint than what can be extracted from the 2d CFT.

\section{The Four Point Function}
\label{GCA4pt}

In this section we make the most nontrivial check, as yet, of the consistency of the 
scaling limit on the 2d CFT. First we consider the GCA differential equation
for the four point function of GCA primaries one of which is $\phi_{2(1,1)}$ 
which has a level two null descendant. The general solution of this equation 
consistent with the crossing symmetry of the problem is discussed. We then proceed to
rederive this result by taking the scaling limit of the corresponding four point function in the 
2d CFT. One finds a solution in the scaling limit which is of the general form inferred from the GCA but
now further constrained by the requirements of monodromy invariance present in the parent CFT. 
Finally, as a zeroth order consistency check of the full theory we briefly examine the factorisation of the four point answer into three point functions and find results in agreement with the fusion rules derived in the previous section.

\subsection{GCA Four Point Function}
\label{GCA4ptfromGCA}
We study the correlation function
of  four GCA primary fields (in the nonrelativistic Kac table)
\begin{equation}
G_{\rm GCA}^{(4)}(\{t_i,x_i\})
=
\langle
\phi_{r_0(s_0,s_0')}(t_0,x_0)
\phi_{r_1(s_1,s_1')}(t_1,x_1)
\phi_{r_2(s_2,s_2')}(t_2,x_2)
\phi_{r_3(s_3,s_3')}(t_3,x_3)
\rangle. \label{G(4)}
\end{equation}
By solving the Ward identities coming from the symmetries 
$L_{0,\pm 1}$, $M_{0,\pm 1}$,
the form of the four point function is restricted to
\begin{equation}
G_{\rm GCA}^{(4)}(\{t_i,x_i\})
=
\prod_{0\leq i < j \leq 3}
t_{ij}^{{1 \over 3} \Sigma_{k=0}^3 \Delta_k - \Delta_i - \Delta_j} 
{\rm e}^{ -{ x_{ij} \over t_{ij} } ({1 \over 3} \Sigma_{k=0}^3 \xi_k - \xi_i - \xi_j)
} {\cal G}_{\rm GCA}(t,x). \label{G(t,x)}
\end{equation}
Here $\Delta_i$ and $\xi_i$
are defined by (\ref{fus1-2}) and (\ref{fus1-3})
with replacing $\{r(s,s')\} \to \{r_i(s_i,s_i')\}$.

The non-relativistic analogues of the 
cross ratio $t$ and $x$, which are defined by
\begin{equation}
t = {t_{01} t_{23} \over t_{03} t_{21}}, \qquad
{x \over t}
=
{x_{01} \over t_{01}}
+
{x_{23} \over t_{23}}
-
{x_{03} \over t_{03}}
-
{x_{21} \over t_{21}},
\label{t,t/x}
\end{equation}
are invariant under the coordinate 
transformation $L_{0,\pm 1}$, $M_{0,\pm 1}$.
Hence the function ${\cal G}_{\rm GCA}(t,x)$ is not 
determined from these symmetries. 
As explained in Sec.\ref{DiffEqGCA},
for null states 
we have differential equations 
which can be used to further 
restrict the four point function.
In the following we consider the 
differential equations coming from
the primary $\{r_0(s_0,s_0')\}=\{2(1,1)\}$.
So, in the following $(\Delta_0,\xi_0) = (\Delta,\xi)$ given in (\ref{fus1}).

The differential equations coming
from the null states $|\chi^{(1)} \rangle$ 
and $|\chi^{(2)} \rangle$ are given by
\begin{align}
&
[\,
{1 \over 2} \partial_{x_0}^2
+
\sum_{i=1}^3 
\kappa^{(-1)} ( {\xi_i \over t_{i0}^2} + {1 \over t_{i0}} \partial_{x_i})\,]
G^{(4)}_{\rm GCA} \label{DG(4)1} =0, \\ 
&[\,
\partial_{x_0}  \partial_{t_0}
-
{1 \over 2}
{\kappa^{(0)} \over \kappa^{(-1)}}
\partial_{x_0}^2
-
\sum_{i=1}^3
\kappa^{(-1)}
(
{2 \xi_i x_{i0} \over t_{i0}^3} 
+ 
{\Delta_i \over t_{i0}^2}
+
{x_{i0} \over t_{i0}^2}\partial_{x_{i}}
-
{1 \over t_{i0}} \partial_{t_{i}}
)\,]
G^{(4)}_{\rm GCA} \label{DG(4)2} = 0,
\end{align}
respectively. Here we have introduced the notation
\begin{equation}
\kappa^{(-1)} = - {2 \over 3} \xi_0,\quad
\kappa^{(0)} = {2 \over 3} \Delta_0 + {2 \over 3}.
\label{kappa}
\end{equation}
By substituting (\ref{G(t,x)}) into (\ref{DG(4)1}) 
and (\ref{DG(4)2}), we obtain  differential equations 
for ${\cal G}_{\rm GCA}(t,x)$.
We can set $t_1=0$, $t_2=1$, $t_3=\infty$ and 
$x_1=x_2=x_3=0$, by using the finite part of GCA 
(we can use the $L_{0,\pm 1}$ to fix the $t$'s
and then the $M_{0,\pm 1}$ to fix the $x$'s),
so that we have $t_0=t$, $x_0=x$. 
Further introducing $H(t,x)$ as 
\begin{equation}
|t|^{ \Sigma_{i=0}^3 {\Delta_i \over 3} - \Delta_0 - \Delta_1}  
|1-t|^{\Sigma_{i=0} {\Delta_i \over 3} - \Delta_0 - \Delta_2}
{\cal G}_{\rm GCA}(t,x) =
{\rm e}^{{x \over t}
(\Sigma_{i=0}^3 {\xi_i \over 3} -\xi_0 - \xi_1)-{x \over 1-t}
(\Sigma_{i=0}^3{\xi_i \over 3} -\xi_0 - \xi_2)}
H(t,x), \notag
\end{equation}
the differential equations for $H(t,x)$ are given by 
\begin{align}
&\partial_x^2 H
+
2\kappa^{(-1)}
{1 - 2 t \over t(1-t)} \partial_x H
+
2\kappa^{(-1)}
\bigg(
{\xi_1 \over t^2}
+
{\xi_2 \over (1-t)^2}
+
{\xi_0+\xi_1+\xi_2-\xi_3 \over t (1-t)}
\bigg)H=0, \\
&\partial_t \partial_x H
-
{1 \over 2} {\kappa^{(0)} \over \kappa^{(-1)}}
\partial_x^2 H
+
\kappa^{(-1)} 
\bigg\{{1 - 2 t \over t(1-t)} \partial_t H 
- x
{1 - 2 t + 2 t^2 \over t^2 (1-t)^2}
\partial_x H
\nonumber \\
& \hspace{2cm}
-
\bigg(
{\Delta_1 \over t^2}
+{\Delta_2 \over (1-t)^2}
+{\Delta_0+\Delta_1+\Delta_2-\Delta_3 \over t(1-t)} 
\bigg)H \nonumber \\
& \hspace{2cm} 
- x 
\bigg( 
2 {\xi_1 \over t^3}
-
2 {\xi_2 \over (1-t)^3}
+
(\xi_0+\xi_1+\xi_2- \xi_3)
{1 - 2t \over t^2 (1-t)^2}
\bigg)H
\bigg\}=0.
\end{align}
The first equation can be easily solved 
to give the following two independent solutions:
\begin{equation}
H_\pm(t,x)
=
(D(t))^{-{1 \over 2}}
|t(1-t)|^{- \kappa^{(0)} + 1}
{\cal H}_\pm (t)
\exp
\bigg\{
{x C_2 \over t (1 - t)}
\big(
- 1 + 2 t \pm \sqrt{D(t)}
\big)
\bigg\},
\label{Hpm(t,x)}
\end{equation}
where ${\cal H}_\pm (t)$ 
are undetermined functions and we have extracted some 
$t$-dependence just for later convenience.
The function $D(t)$ is defined by
\begin{equation}
D(t) = r_1^2 (1-t) + r_2^2t - r_3^2 t (1-t),
\label{D(t)}
\end{equation}
and it satisfies $D(t) \geq 0$
because of the triangle inequality 
for $(r_1,r_2,r_3)$, which follows 
from the fusion rule (see appendix \ref{FusionRuleAnd}). 

Now in terms of ${\cal H}_\pm (t)$, 
the second equation for $H(t,x)$ is simplified as
\begin{equation}
\partial_t \log {\cal H}_\pm
=
\pm
{1 \over \sqrt{D}}
\bigg\{
{{\cal C}_1 \over t}
+
{{\cal C}_2 \over 1-t}
-
{{\cal C}_3}
\bigg\}.
\label{dlogH}
\end{equation}
Here $t$-independent constants ${\cal C}_i$ are 
defined by
\begin{eqnarray}
{\cal C}_i 
& = &
{1 \over 3} r_i^2 \Delta_0 
+ {1 \over 3} r_i^2 
+ \Delta_i 
+ {1 \over 3} \Delta_0
- {2 \over 3} \\
& = &
\pm r_i ( \Delta_{r_i \pm 1(s_i,s_i')} 
- \Delta_0 - \Delta_i + \kappa^{(0)} - 1).
\label{calC}
\end{eqnarray}
The second line is quite a remarkable 
simplification.
In fact it has a simple physical interpretation
with $\Delta_{r_i \pm 1(s_i,s_i')}$ corresponding to 
the conformal dimension of the allowed intermediate states.
We will return to this point later in Sec. \ref{factorization}.

Finally the differential equation (\ref{dlogH})
can be solved if we notice the relation
\begin{equation}
\partial_t \log|r_1 + r_3 t + \sqrt D| 
=
{1 \over 2 \sqrt D} {1 \over t} (- r_1 + r_3 t + \sqrt D).
\end{equation}
Notice that this equation still holds after 
a flip of the signs in front of $r_1$ and/or $r_3$,
and it also gives an additional relation by replacing 
$r_1 \leftrightarrow r_2$ and $t \to 1-t$.
By taking linear combinations of these equations,
we obtain the solutions of the above differential 
equations which are given by
\begin{equation}
{\cal H}_+ (t)  = {\cal I}_{1,2,3}(t),\quad 
{\cal H}_- (t)  = {\cal I}_{2,1,3}(1-t),
\label{H+,H-}
\end{equation}
with
\begin{eqnarray}
{\cal I}_{1,2,3}(t) & = &
\bigg|
{r_1 + r_2 + r_3 \over r_1 + r_2 - r_3 }
{- r_1 + r_3 t + \sqrt D \over  r_1 + r_3 t + \sqrt D}
\bigg|^{{{\cal C}_1 \over r_1}} 
\bigg| 
{r_1 + r_2 + r_3 \over r_1 + r_2 - r_3}
{r_2 - r_3 (1-t) + \sqrt D \over r_2 + r_3 (1-t) - \sqrt D}
\bigg|^{{{\cal C}_2 \over r_2}}
\nonumber \\
&& \hspace{3cm}
\times \bigg|
{r_1+r_2-r_3 \over -r_1 + r_2 + r_3}
{r_2 + r_3 (1-t) + \sqrt D 
\over 
r_2 - r_3 (1-t) + \sqrt D}
\bigg|^{{{\cal C}_3 \over r_3}}. \label{I}
\end{eqnarray}
The overall constant factor is chosen 
for later convenience.

Thus the four point function is given by a general
linear combination of the two solutions $H_\pm(t,x)$,
which are defined by (\ref{Hpm(t,x)})
with the functions (\ref{H+,H-}).
In particular, we may allow different linear combinations 
for the different segments $t<0$, $0<t<1$ and $1<t$,
since the differential equations are solved 
independently for each segment.
It may be worth pointing out that the term in each 
absolute value sign takes a definite signature 
for the each segment.
In the relativistic 2d CFT, we would determine 
the particular linear combination by requiring 
the four point function to be singlevalued on the 
complex $z$ plane.
In the case of the GCA such an argument is not available.
Thus we will discuss only the constraint coming from  
crossing symmetry.

First of all, it is clear that the following linear combination 
is invariant under the exchange $1 \leftrightarrow 2$ 
and $(t,x) \leftrightarrow (1-t, -x)$:
\begin{equation}
H(t,x) = H_+ (t,x) + H_- (t,x).
\label{H=H++H-}
\end{equation}
Next, by using the following property
of the function ${\cal I}_{1,2,3}(t)$:
\begin{align}
{\cal I}_{3,2,1}(1/t) & =
\begin{cases}
{\cal I}_{1,2,3}(t)\,, \\ 
{\cal I}_{2,1,3}(1-t) \,,
\end{cases} 
{\cal I}_{2,3,1}(1-1/t) =
\begin{cases}
{\cal I}_{2,1,3}(1-t)\,, & \quad  (t > 0)\,, \\
{\cal I}_{1,2,3}(t)\,, & \quad (t<0)\,,
\end{cases}
\end{align}
it is easy to show that $H(t,x)$ behaves 
correctly under the exchange $1 \leftrightarrow 3$ 
and $(t,x) \leftrightarrow (1/t,- x/t^2)$, namely,
\begin{equation}
H(t,x) \to |t|^{ 2 \Delta_0} {\rm e}^{ - 2 \xi_0{x \over t}} H(t,x)\,.
\end{equation}
This would be the GCA analogue of the 
transformation 
$G(z,\bar z) \to z^{2h} \bar z^{2 \bar h} G(z,\bar z)$
in 2d CFT.

Then we may ask whether (\ref{H=H++H-}) is the unique 
linear combination which has these properties.
In fact since we should allow independent coefficients 
for different segments 
(see footnote \ref{relativesign}),
we have infinitely many 
combinations which have the correct property 
of crossing symmetry.
Here we will not pursuit the most general
form of such possibilities, but just 
write down a class of such combinations:
\begin{equation}
H(t,x) = 
\begin{cases}
f_{123} H_+(t,x) + f_{321} H_-(t,x), & t < 0, \\
f_{132} H_+(t,x) + f_{231} H_-(t,x), & 0 < t < 1, \\
f_{312} H_+(t,x) + f_{213} H_-(t,x), & 1 < t.
\end{cases}
\label{4ptGCAfinal}
\end{equation}
Here $f_{123}$ is an arbitrary function of 
the quantum numbers of the three fields 
$\phi_{r_i(s_i,s_i')}$ ($i=1,2,3$).
It is obvious that this combination has 
the same property as (\ref{H=H++H-}).
Also, by changing the relative sign 
of $H_+$ and $H_-$, we would obtain 
the combination which flips the overall
sign under the exchange 
$1 \leftrightarrow 2$, 
$(t,x) \leftrightarrow( 1-t, -x)$
or/and 
$1 \leftrightarrow 3$, 
$(t,x) \leftrightarrow(1/t, -x/t^2)$.
We see no reason to forbid these combinations.

Thus, we find that the requirement of crossing symmetry
does not fix the GCA four point function uniquely.
In the next section, we consider the corresponding  four point
function in the 2d CFT and take the non-relativistic limit.
In this case, the original four point function is fixed uniquely
by the requirement of singlevaluedness.
It would be interesting to study whether further conditions
coming directly from GCA context would 
fix the four point function completely or not.
We will leave this issue as a problem for the future.

\subsection{GCA Four Point Function from 2d CFT}
\label{CFT4pttoGCA4pt}
Next we discuss the four point function 
starting from the 2d CFT and taking the limit. 
We take the holomorphic part of the
four point function in 2d CFT as\footnote{
Hereafter, to avoid complexity,
we use the notation such as
$\phi_i$ and $h_i$ 
for the objects corresponding to the member 
$(r_i,s_i)$ on the 2d CFT Kac table.
}
\begin{equation}
G_{\rm 2d\,CFT}^{(4)}(\{z_i\}) =
\langle
\phi_0(z_0)
\phi_1(z_1)
\phi_2(z_2)
\phi_3(z_3)
\rangle  =
\prod_{0 \leq i<j \leq 3} 
z_{ij}^{{1 \over 3}\sum_{k=0}^3 h_k - h_i - h_j} 
{\cal G}_{\rm 2d\,CFT}(z),
\label{2dCFT4pt}
\end{equation}
where the cross ratio is defined by
$z = (z_{01}z_{23})/(z_{03} z_{21})$.
By taking $(r_0,s_0)=(2,1)$,
we obtain the differential equation of the form 
\begin{equation}
(\hat {\cal L}_{-2} + \eta \hat {\cal L}_{-1}^2) 
G_{\rm 2d\,CFT}^{(4)}(\{ z_i \})
=
0,
\end{equation}
where $\eta$ is introduced in (\ref{nullcond}) and 
the differential operators $\hat {\cal L}_{-n}$ 
are defined in \eqref{calL(-k)}.

We introduce the function $K(z)$ as
\begin{align}
z^{\beta_1}(1-z)^{\beta_2}K(z)
& =
z^{\Sigma_{i=0}^3 {h_i \over 3} - h_0 - h_1}
(1-z)^{\Sigma_{i=0}^3 {h_i \over 3} -h_0 - h_2}
{\cal G}_{\rm 2d\,CFT}(z) \label{K=calG}\\[1mm]
&
= 
\lim_{z_3 \to \infty} z_3^{2 h_3} 
G_{\rm 2d\,CFT}^{(4)}(z_0 = z,z_1=0,z_2=1,z_3)\,,
\label{limG}
\end{align}
with
$\beta_i$ defined as any one of the solutions of the quadratic equation
\begin{equation}
\beta_i (\beta_i - 1) + \kappa \beta_i - \kappa h_i = 0, 
\qquad 
(\kappa = - \eta^{-1}).
\label{beta_i}
\end{equation}
The second expression (\ref{limG}) is upto 
a constant phase.
Then the differential equation is rewritten into the 
standard form of the Hypergeometric differential equation:
\begin{equation}
z(1-z) \partial_z^2 K(z) 
+ 
(\gamma - (\alpha + \beta + 1) z ) \partial_z K(z)
-
\alpha \beta K(z) = 0.
\label{DiffHyper}
\end{equation}
The parameters $\alpha$, $\beta$ and $\gamma$
are given by
\begin{eqnarray}
&& \gamma  =  
2 \beta_1 + \kappa,  \quad 
\alpha + \beta + 1  =  2 (\beta_1 + \beta_2 + \kappa), 
\label{gamma,alpha,beta(1)}\\
&& \alpha \beta  =  
\beta_1(\beta_1 - 1) 
+ 2 \beta_1 \beta_2 
+ \beta_2(\beta_2 - 1) 
+
\kappa( 2 \beta_1 + 2 \beta_2 + h_0 - h_3).
\label{gamma,alpha,beta(2)}
\end{eqnarray}
The two independent solutions of (\ref{DiffHyper}) 
which diagonalize
the Monodromy transformation $z \to {\rm e}^{2 \pi i} z$
around $z=0$ are given by 
$K_1(\alpha,\beta,\gamma,z) = F(\alpha,\beta,\gamma,z)$ and 
$K_2(\alpha,\beta,\gamma,z) 
= z^{1-\gamma} F(\alpha-\gamma+1,\beta-\gamma+1,2-\gamma,z)$,
where $F(\alpha,\beta,\gamma,z)$ is defined as 
the Hypergeometric series and its analytic 
continuation.
The argument so far would be applied in a parallel way
to the anti-holomorphic part with the replacement of
$h_i \to \bar h_i$, $\beta_i \to \bar \beta_i$,
$\kappa \to \bar \kappa$ and 
$(\alpha,\beta,\gamma) \to (\bar \alpha, \bar \beta, \bar \gamma)$,
along with the replacement of the coordinate $z \to \bar z$.

Now we construct the full four point function 
by multiplying the holomorphic and the anti-holomorphic
contributions and taking an appropriate linear combination. 
A standard argument (see for e.g. \cite{Ketov:1995yd}) 
for choosing the combination 
is based on the singlevaluedness of the 
full four point function on the complex plane.
In fact, by considering the behaviour under the 
Monodromy transformations $z \to {\rm e}^{2 \pi i}z$
($1-z \to {\rm e}^{2 \pi i}(1-z)$) around 
$z=0$ ($z=1$), we find that the following 
combination:
\begin{equation}
I(z,\bar z) = 
K_1(\alpha,\beta,\gamma,z)
K_1(\bar \alpha,\bar \beta,\bar \gamma,\bar z) 
+
A
K_2(\alpha,\beta,\gamma,z)
K_2(\bar \alpha,\bar \beta,\bar \gamma, \bar z)\,,
\label{M-Invform}
\end{equation}
with
\begin{equation}
A = - 
{
\Gamma(\gamma) \Gamma(1-\beta)\Gamma(1-\alpha)
\over
\Gamma(2 - \gamma) \Gamma(\gamma - \alpha) \Gamma(\gamma - \beta)
}
{
\Gamma(\bar \gamma) 
\Gamma(\bar \alpha - \bar \gamma + 1)
\Gamma(\bar \beta - \bar \gamma + 1)
\over 
\Gamma(\bar \alpha) \Gamma(\bar \beta)
\Gamma(2 - \bar \gamma)
},
\label{A}
\end{equation}
is the unique singlevalued combination, provided  
$\alpha - \bar \alpha$, $\beta - \bar \beta$
and $\gamma - \bar \gamma$ are all integers.
To see this one may use the formula (\ref{FtoF+F(2)}).

Next we take the non-relativistic limit.
For the purpose, we use the asymptotic form
of the Hypergeometric functions for large values 
of $\alpha$, $\beta$ and $\gamma$, which can 
be derived by the saddle point analysis 
of the integral formula:
\begin{eqnarray}
F(\alpha,\beta,\gamma,z)
& = & 
{\Gamma(\gamma) \over \Gamma(\beta) \Gamma(\gamma-\beta)}
\int_0^1 dw 
w^{\beta - 1} 
(1 - w)^{\gamma - \beta -1}
(1 - z w)^{-\alpha},
\label{IntHyper}
\end{eqnarray}
where the conditions $\gamma > \beta > 0$ 
and that $z$ does not take a real value greater than 
$1$ are assumed. 
We expand every parameter as 
\begin{align}
\beta_i
=
{1\over \epsilon}\beta_i^{(-1)} + \beta_i^{(0)} + \cdots, \quad 
\alpha
= 
{1 \over \epsilon} \alpha^{(-1)} + \alpha^{(0)} + \cdots, \quad
{\rm etc.}
\end{align}
Then, the quadratic equations for $\beta_i$'s 
are solved by 
\begin{equation}
\beta_i^{(-1)} = - {1 \over 2} C_2(1 + \delta_i r_i),\quad 
\beta_i^{(0)} =  
- {1 \over 2}\delta_i {\ell}_i - {1 \over 2}( \kappa^{(0)} - 1),
\quad (\delta_i = \pm 1 ),
\label{main-betai}
\end{equation}
with
\begin{equation}
\ell_i =
{
(\kappa^{(0)}-1) \kappa^{(-1)} 
+ 
2 \kappa^{(-1)} h_i^{(0)}
+
2 \kappa^{(0)} h_i^{(-1)}
\over C_2 r_i
}
=
\Big({13 \over 6} - C_1 \Big) r_i - s_i.
\label{li}
\end{equation}
Since $\beta_i$ are auxiliary parameters
introduced in order to change the form of the differential
equation, we take 
$(\delta_1,\delta_2) = (-1,+1) $ 
without loss of generality.
Also, since the Hypergeometric function
is symmetric under the exchange
$\alpha \leftrightarrow \beta$,
we take $\alpha^{(-1)} < \beta^{(-1)}$.
Then the solutions
$(\alpha,\beta,\gamma)$ 
of Eqs.(\ref{gamma,alpha,beta(1)})
(\ref{gamma,alpha,beta(2)})
are expanded as
\begin{equation}
\begin{array}{lll}
\displaystyle
\alpha^{(-1)}
=
{C_2 \over 2}(r_1 - r_2 - r_3),\quad &
\displaystyle \beta^{(-1)}
= {C_2 \over 2} (r_1 - r_2 + r_3), \quad &
\displaystyle
 \gamma^{(-1)} = C_2 r_1,\\[3mm]
\displaystyle 
\alpha^{(0)} 
= {1 \over 2}(\ell_1 - \ell_2 - \ell_3 + 1), \quad & 
\displaystyle
\beta^{(0)} =
{1 \over 2} (\ell_1 - \ell_2 + \ell_3 + 1), \quad & 
\gamma^{(0)} = \ell_1 + 1,
\end{array} 
\label{abcexpansion}
\end{equation}
where $\ell_3$ is defined by (\ref{li}).
As for the expansion of the parameters 
of the anti-holomorphic part, we first 
choose the leading term of $\bar \beta_i$ 
as $\bar \beta_i^{(-1)} = - \beta_i^{(-1)}$.
This can be achieved by taking appropriate
branch of the quadratic equation for 
$\bar \beta_i$. 
Then the leading order terms of the
rest of the parameters are given just by flipping 
the sign as 
$(\bar \alpha^{(-1)},\bar \beta^{(-1)},\bar \gamma^{(-1)})
=-
( \alpha^{(-1)}, \beta^{(-1)},\gamma^{(-1)})$,
while ${\cal O}(1)$ term is independent from those of the 
holomorphic part, since we consider different $s$ and $s'$ as
(\ref{hh1}) and (\ref{hh2}).
So, we introduce the notation $\bar \ell_i $,
which are defined by (\ref{li}) 
with changing $s_i \to s'_i$.
Then the ${\cal O}(1)$ terms are given by
replacing $\ell_i$ with $\bar \ell_i$
as $\bar \gamma^{(0)} = \bar \ell_1+1$, and 
the same for $\bar \alpha^{(0)}$ and $\bar \beta^{(0)}$.

For this parameter choice, 
the integral (\ref{IntHyper}) can be evaluated 
by taking account of a single saddle point located 
on the segment $0 < w < 1$, and the result is given by 
\begin{equation}
K_1(\alpha,\beta,\gamma,z)
\to 
\bigg(
{\epsilon \over 2 \pi}
\bigg)^{1 \over 2}
{\Gamma(\gamma) \Gamma(1-\alpha) \over \Gamma(\gamma - \alpha)}
K_{+}(z), \quad (z = t + \epsilon x, \quad t < 1),
\label{asymptoticF}
\end{equation}
where the function $K_+$ 
is defined by (\ref{calK(1)}) and (\ref{calK(2)})
($\delta_{C_2}$ in (\ref{calK(1)}) is the
sign of $C_2$, which we take $+1$ in the main text).
See the appendix \ref{AsymHyper} for 
a detailed discussion on the 
saddle point analysis,
where the asymptotic form of the rest 
of the region $1 < t$, the other solution 
$K_2=z^{1-\gamma} F(\alpha-\gamma+1,\beta-\gamma+1,2-\gamma,z)$,
and the anti-holomorphic counterparts of them are also discussed.

In particular, for the region $t < 0$,
the asymptotic form of the basis functions
are given by Eqs.(\ref{zF})--(\ref{bzF}) 
in terms of the functions $K_\pm (z)$ 
and $\bar K_\pm (\bar z)$.
Inserting these into (\ref{M-Invform}),
it is easy to show that the ``cross term''
$K_+(z) \times \bar K_-(\bar z)$ cancels and 
we obtain the following asymptotic 
form of the function $I(z,\bar z)$
for $z=t + \epsilon x$ and $\bar z = t - \epsilon x$:
\begin{align}
&I(z,\bar z) 
\to
K_+ (z)\bar K_+ (\bar z) \notag \\
&\quad +
(-1)^{n_\beta + n_\gamma}
z^{-{C_2 \over \epsilon}  r_1 - \ell_1}
\bar z^{{C_2 \over \epsilon} r_1 - \bar \ell_1}
(1-z)^{{C_2 \over \epsilon} r_2 + \ell_2}
(1-\bar z)^{- {C_2 \over \epsilon} r_2 + \bar \ell_2}
K_-(z) \bar K_-(\bar z)\,.
\label{asymptI}
\end{align}
Here we have omitted the overall $(t,x)$-independent constant
and $n_\gamma = \bar \gamma - \gamma$,
$n_\beta = \bar \beta - \beta$ are 
integers (see the next subsection).
The asymptotic form of the other regions 
are also derived in the same way and 
we obtain the same result (\ref{asymptI})
(See appendix \ref{AsymHyper}).

Now remember that the 
four point function is given 
by multiplying the additional factor 
$z^{\beta_1}(1-z)^{\beta_2}$ of 
(\ref{K=calG}) (and also the anti-holomorphic factor)
with $I(z,\bar z)$.
By further expanding with respect to $\epsilon$
which appears in $z= t + \epsilon x$ and $\bar z = t - \epsilon x$,
all singular exponents cancel
among the holomorphic and the anti-holomorphic 
contribution.
Finally we obtain the following form of the
four point function:\footnote{We have used 
the assumption that $C_2/\epsilon$ is an even integer
(see footnote \ref{ModularInv}).}
\begin{align}
H(t,x) =
\begin{cases}
(-1)^{n_1} H_+(t,x) + (-1)^{n_2+n_3} H_-(t,x)\,,
& t < 0\,, \\
\phantom{(-1)^{n_1}}H_+(t,x) + (-1)^{n_1 + n_2 + n_3} H_-(t,x)\,,
& 0 < t < 1\,, \\
(-1)^{n_2} H_+(t,x) + (-1)^{n_1+n_3} H_-(t,x)\,,
& 1 < t \,.
\label{4ptfinal}
\end{cases}
\end{align}
Here the functions $H_\pm (t,x)$ are 
the ones introduced in the previous subsection and 
$n_i$ are defined by
\begin{equation}
n_i = {C_2 \over \epsilon} r_i + {1 \over 2} (s_i' - s_i).
\end{equation}
As we will explain in the next subsection, 
the differences $s_i' - s_i$ are now even integers.
Hence the factors like $(-1)^{n_1}$
in (\ref{4ptfinal}) are just signs.

The behaviour of the function (\ref{4ptfinal})
under the exchange $1 \leftrightarrow 2$
and $1 \leftrightarrow 3$ can be easily found as
\begin{align}
H(t,x) \to
\begin{cases}
(-1)^{n_1+n_2+n_3} H(t,x)\,, & 
(1 \leftrightarrow 2, (t,x) \to (1-t,-x))\,, \\
(-1)^{n_2} |t|^{2 \Delta_0} {\rm e}^{-2 \xi_0 {x \over t}} H(t,x)\,, & 
(1 \leftrightarrow 3, (t,x) \to (1/t,-x/t^2))\,.
\end{cases}
\end{align}
Here we should notice that the above monodromy
argument does not fix the overall factor
$C_{1,2,3}^{(4)}$ of the four point function,
which should depend on the quantum numbers 
of the fields $\phi_{i}$ ($i=1,2,3$).
For example, if it satisfies the conditions 
$C^{(4)}_{2,1,3}= (-1)^{n_1+n_2 + n_3}C^{(4)}_{1,2,3}$
and 
$C^{(4)}_{3,2,1}= (-1)^{n_2}C^{(4)}_{1,2,3}$\,,
then the four point function including the
overall constant belongs to the class which
is written down in (\ref{4ptGCAfinal}), with
$f_{123} =  (-1)^{n_1}C_{1,2,3}^{(4)}$.
Deriving the explicit form of the overall
constant requires the three point functions of the theory,
whose overall factor we have also not determined in
the present paper.

In summary, contrary to the discussion 
in the previous section, the asymptotic form
of the four point function (\ref{4ptfinal})
is determined upto an overall constant factor.

\subsection{Singlevaluedness Condition}
\label{Sec:Singlevalued}
Now we study the issue of the singlevaluedness in detail.
It contains two aspects. 
As we mentioned below the equation (\ref{A}), 
the singlevaluedness of the 
function $I(z,\bar z)$ requires $\bar \alpha - \alpha$,
$\bar \beta - \beta$ and $\bar \gamma - \gamma$
to be integers.
Another condition comes from the 
singlevaluedness of the factor 
$z^{\beta_1} \bar z^{\bar \beta_1}
(1-z)^{\beta_2}(1-\bar z)^{\bar \beta_2}$,
which requires $\bar \beta_1 - \beta_1$ and 
$\bar \beta_2 - \beta_2$ to be integers.

Before studying these conditions,
let us mention the singlevaluedness
of the two point functions
of the members on the Kac table, 
which is given by
\begin{equation}
h_{r,s}- \bar h_{r,s'} = {1 \over 2} \times {\rm integer}\,.
\end{equation}
From (\ref{hh1}) and (\ref{hh2}), we have
\begin{equation}
h_{r,s} - \bar h_{r,s'} 
= 
- {1 \over 2} {C_2 \over \epsilon} ( 1- r^2 ) 
- {1 \over  2} r(s - s') + {\cal O}(\epsilon)\,.
\end{equation}
Then the above condition requires that 
$-{C_2 \over 2\epsilon }(1 - r^2) + {\cal O}(\epsilon)$
is an integer or a half integer.
Although it is still 
not very clear in this parameterization,
more detailed study\footnote{
If we parameterize as $c = 13 - 6 (\kappa + 1/\kappa )$ and 
$\bar c = 13 - 6 ( \bar \kappa + 1/ \bar \kappa)$ with $\kappa \gg 1$
and ${\bar \kappa} \ll - 1$, then the 
singlevaluedness requires both $\kappa-\bar \kappa$ and 
$1/\kappa - 1/ \bar \kappa$ to be integers, which is satisfied 
only in the case with $1/\kappa=1/\bar \kappa =0$.} 
tells us that it is satisfied only
in the strict limit $\epsilon = 0$ with 
a ``large'' integer $C_2/\epsilon $.
This final point is always satisfied 
in the modular invariant 2d CFT, in which
case $C_2/\epsilon$ is required to be an even
integer (see the footnote \ref{ModularInv}).

Now from (\ref{li}) and (\ref{abcexpansion}),
we first notice that $\bar \gamma - \gamma$ 
is an integer because of the same reason as above.
Then in order for the differences $\bar \beta - \beta$ 
and $\bar \alpha - \alpha$ to be integers, 
$\Sigma_{i=1}^3 (s_i' - s_i)$ must be an even integer.
On the other hand the requirement for 
$\bar \beta_i - \beta_i$ to be integers
give rise to the condition that 
each $s_i' - s_i$ ($i=1,2$) is an even integer.
In fact, the latter condition is directly related to the 
singlevaluedness of the three point function 
which includes the primary field $\phi_{2(1,1)}$,
namely, the single valuedness of the
three point function 
$\langle \phi_{2(1,1)} \phi_{r(s,s')} \phi_{r\pm 1(s,s')}\rangle$
gives rise to the condition that $s' - s$ is an even integer.\footnote{
It can be shown explicitly by using (\ref{hh1}) and (\ref{hh2})
and requiring that the summation of the spins of the three 
fields is an integer.}
This just means that the primary field $\phi_{2(1,1)}$
interacts only with the primary field having an
even $s' - s$. For an odd $s' - s$, 
the three point function, and hence the 
four point function, vanishes.
Then for the non-vanishing four point function, 
$s_i' - s_i$ ($i=1,2,3$) are automatically even integers 
and the singlevaluedness condition is satisfied.
A relation between the parameter 
$\beta_i$ and 
the three point function is further
clarified in the next section.

\subsection{A Quick Look at Factorization and the Fusion Rule}
\label{factorization}
In 2d CFT, the four point function 
is written as the summation over 
the various intermediate fields.
Let us consider the four point function 
(\ref{2dCFT4pt}), with its anti-holomorphic part,
and set $z_0 = z$, $z_1=0$, $z_2=1$ and $z_3=\infty$. 
Then for small $z$, the four point function
is given by 
\begin{eqnarray}
\sum_{p,\{\vec k ,\vec {\bar k} \}}
C_{01}^{p,\{\vec k, \vec \bar k\}} 
z^{h_p-h_0-h_1+K} 
\bar z^{\bar h_p-\bar h_0-\bar h_1+\bar K}
\langle 
\phi_p^{\{\vec k, \vec {\bar k}\}}(0)
\phi_2(1) \phi_3(\infty)
\rangle. \label{4ptOPE}
\end{eqnarray}
Here $\phi_p^{\{\vec k, \vec {\bar k}\}}$ 
is the intermediate primary field (for $\vec k = \vec {\bar k} = 0$),
and its descendants (for nontrivial $\vec k$ and $ \vec {\bar k}$),
which are given by acting, on the primary,
series of the operators 
${\cal L}_{-k_n} \cdots {\cal L}_{-k_1}$ and 
the same for the anti-holomorphic part.
These intermediate states appear in the 
OPE of $\phi_0$ and $\phi_1$.
This relation shows that the four point function
is essentially written in terms of the three point functions.
It would be interesting to study the 
similar property in detail in the case of the GCA.
Here, as a first step in such an analysis, 
we study the contribution coming from the 
intermediate primary fields. 

In the previous subsections, we studied the 
four point function of the primary fields
with setting one of the field to have
$\{r_0(s_0,s_0')\}=\{2(1,1)\}$
on the Kac table.
In this case, the possible primary fields
$\phi_p^{\{0,0\}}$ appearing in \eqref{4ptOPE} 
are the ones with $r_p(s_p,s_p') = r_1\pm 1(s_1,s_1')$.
Let us see this from the equation (\ref{beta_i}).
Remember that the holomorphic part of
the four point function is given 
in terms of the Hypergeometric functions
$K_1(z)$ and $K_2(z)$ as 
\begin{equation}
\lim_{z_3 \to \infty} z_3^{2 h_3} G^{(4)}_{\rm 2d\,CFT}(\{z_i\}) 
= 
z^{\beta_1} (1-z)^{\beta_2}
\big( a_1 K_1(z) + a_2 K_2(z) \big),
\end{equation}
where $a_i$ are coefficients.
Hence for small $z$, we have two channels
contributing to the four point function, namely,
one is $z^{\beta_1} K_1(z) \sim z^{\beta_1}$
and the other is 
$z^{\beta_1} K_2(z) \sim z^{\beta_1 -\gamma+1} 
= z^{-\beta_1 - \kappa + 1}$.
In terms of the intermediate states of (\ref{4ptOPE}),
this means there are two intermediate conformal families.
The conformal dimensions of the primary fields of these families 
are given by $h_{p_+} = h_0+h_1 + \beta_1$ and 
$h_{p_-} = h_0+h_1- \beta_1 - \kappa +1$.
Now we may notice that these two exponents 
$\beta_1$ and $- \beta_1 - \kappa +1$ 
are the two solutions of (\ref{beta_i}),
which means that $h_{p_{\pm}}$ are 
indeed the conformal dimensions which 
are allowed from the relativistic version
of the fusion rule, namely, $h_{p_\pm} = h_{r_1 \pm 1}$ 
(or $h_{p_\pm} = h_{r_1 \mp 1}$ depending on the choice
of the branch of the quadratic equation for $\beta_i$).\footnote{
In 2d CFT, the differential equations
for the three point function
$\langle \phi_{2,(1,1)}
\phi_{r_1(s_1,s_1')}
\phi_{r_p(s_p,s_p')} \rangle$
give rise to the following constraint:
$2 ( 2 h_{2,1} + 1 )
(h_{2,1} + 2 h_{r_p,s_p} - h_{r_1,s_1}) =
3 ( h_{2,1} - h_{r_1,s_1} + h_{r_p,s_p}) 
(h_{2,1}-h_{r_1,s_1}+h_{r_p,s_p}+1)$
and the same for the anti-holomorphic part.
This is the same as equation (\ref{beta_i}).}

Next let us study the GCA four point function.
In subsection \ref{GCA4ptfromGCA}, we have seen that the 
GCA four point function is composed of, again,
two solutions of the differential equation,
which we call $H_\pm(t,x)$.
The solution is given by (\ref{Hpm(t,x)}),
with ${\cal H}_{\pm}(t)$ satisfying the
differential equation (\ref{dlogH}).
In fact, the small $t$ behaviour of the four 
point function can be derived by using
Eqs.(\ref{Hpm(t,x)})--(\ref{dlogH}) as 
\begin{eqnarray}
H_\pm(t,x) & \sim &
t^{-\kappa^{(0)}+1} 
\cdot t^{\pm{{\cal C}_1 \over r_1} }
\cdot 
\exp \Big( {x \over t} C_2 (- 1 \pm r_1) \Big) \notag \\
& = &
t^{\Delta_{r_1\pm 1 (s_1,s_1')} - \Delta_0 - \Delta_1}
\exp\Big( -(\xi_{r_1 \pm 1} - \xi_0 - \xi_1 ) {x \over t } \Big).
\end{eqnarray}
These are indeed the two leading behaviours which appear 
for the operator product expansion of the 
primary fields $\phi_{2(1,1)}$ and $\phi_{r_1(s_1,s_1')}$ in GCA
(see appendix \ref{Descendants} for a 
preliminary examination of the GCA OPE).
Now it is clear from (\ref{dlogH}) and (\ref{calC}), that 
the other ordering, i.e., first taking the OPE between 
$\phi_0(z)$ and $\phi_2(1)$, gives two intermediate states
which are again allowed from the fusion rule. 

\section{Concluding Remarks}

This concludes our present study of the quantum mechanical realisation of the 
GCA in two dimensions. 
We found that 2d GCFT's with nonzero central charges $C_1,C_2$ can be
readily obtained by considering a somewhat unusual limit of a
non-unitary 2d CFT.  While the resulting Hilbert space of the GCFT  is
again non-unitary, the theory seems to be otherwise well defined. We
found that many of the structures parallel those in the Virasoro algebra
and indeed arise from them when we realise the GCA by means of the
scaling limit. But in most cases we could also obtain many of the same
results autonomously from the definition of the GCA itself, showing that
these are features of any realisation of this symmetry. Along these
lines, one of the nice things to independently compute would be the form
of the GCA Kac Table which we obtained in Sec. 5 from the scaling limit
of the relativistic table. 
It would be interesting if one can find an analogue of minimal models
for the GCA for which there are a finite number of GCA primaries and for
which the solution of the theory can be completely given. Though  we
must hasten to remind the reader that such theories (if they at all
exist) would not, in any sense, arise from the Virasoro minimal models
which have $0<c<1$.
Another interesting direction would be to investigate possible
generalizations of our results to higher dimensions where 
we do have the GCA but no analogue of the Virasoro Algebra.
In this context it must be clear from the explicit solutions in Sec. 8
that the lack of holomorphicity, while it makes the expressions more
cumbersome, is not a real barrier to solving the theory provided one has
GCA null vectors. 

As far as the scaling limit of the 2d CFT is concerned, it is important to 
make some further checks on its consistency with the requirements from the GCA.
From the point of view of the 2d CFT it would be nice to get a better understanding of 
this scaling limit on the central charges. Sending their magnitude to infinity seems 
like some kind of large $N$ or classical limit. While it is somewhat dismaying that 
the theory is non-unitary, could there perhaps be an interesting "physical" subsector 
which is unitary? It is amusing to note in this context that in perturbative string theories we do have 
2d CFTs with large (26 or even 15!) positive and negative values of $c$
and ${\bar c}$ in the matter and ghost sectors.

In this paper we have only studied the limit of generic 2d CFTs. It
would be interesting to carry out a similar scaling limit on CFT's with
supersymmetry and/or Kac-Moody symmetries.{\footnote{Supersymmetric
versions of the GCA have been recently studied in
\cite{Bagchi:2009ke,deAzcarraga:2009ch,Sakaguchi:2009de}}} More generally, it is important to understand whether there are concrete
physical situations where such a scaling limit is actually
realised. That would make the further study of these symmetries all the
more exciting.

\subsection*{Acknowledgements}

We are happy to acknowledge useful conversations with Vijay Balasubramanian, Shamik Banerjee,
Bobby Ezhuthachan, Matthias Gaberdiel, Jaume Gomis, Ashoke Sen and
specially Shiraz Minwalla for stimulating discussions as well as
comments on the manuscript. The work of R. G. was supported in part by
the Swarnajayanthi fellowship of the Dept.~of Science
and Technology of the Govt.~of India. 
A.B. would like to thank the string theory groups at University of Pennsylvania, MIT, Santa Barbara, Stanford, Chicago, Michigan and the Perimeter Institute where initial results of this work were presented.  All of us would like to record our gratefulness for the support to basic scientific research extended by the people of India. 

\section*{Appendix}
\appendix

\section{Descendants and GCA Conformal Blocks}

\label{Descendants}

\subsection{GCA Descendants}

By means of the differential operators
$\hat L_{-k}, \hat M_{-k}$ (with $k>0$) in \eq{Mm}, 
we may express the correlation function 
including a general GCA descendant with the 
correlation function of its primary
$\phi_{\Delta\xi}(t,x)$. 
We had in fact already used this in Sec.~6 for the 
simple cases of these descendants corresponding to null states
as \eq{nulleq1} and \eq{nulleq2}. 
The general expression can be written as  
\ben{ope1}
&&\langle 
\phi_{\Delta\xi}^{\lbrace  {\vec{k}}, {\vec{q}} \rbrace}
(t,x) 
\phi_1(t_1,x_1) \cdots \phi_n (t_n,x_n)
\rangle \notag \\[1mm]
&& \qquad = {\hat L}_{-k_i} \cdots {\hat L}_{-k_1} 
{\hat M}_{-q_j}\cdots {\hat M}_{-q_1}
\langle
\phi_{\Delta\xi}(t,x) 
\phi_1(t_1,x_1)
\cdots
\phi_n(t_n,x_n)
\rangle\,,
\een
where ${\vec{k}} = (k_1, k_2, \cdots ,k_i) $ 
and ${\vec{q}} = (q_1, q_2, \cdots ,q_j)$
are sequences of positive integers such that 
$k_1\leq k_2 \cdots \leq
k_i$ and similarly for the $q$'s.  
Also note that $\phi_{\Delta\xi}^{\lbrace 0,0 \rbrace}(t,x)$ 
denotes the primary $\phi_{\Delta\xi}(t,x)$ itself.


\subsection{The OPE and GCA Blocks}

Just as in the relativistic case, 
the OPE of two GCA primaries 
can be expressed in terms of 
the GCA primaries and their descendants  as
\be{ope}
\phi_1(t,x) \phi_2(0,0) 
= \sum_{p}  \sum_{\lbrace {\vec{k}} , {\vec{q}} \rbrace}
C_{12}^{p\lbrace  {\vec{k}}, {\vec{q}} \rbrace}(t,x)
\phi_p^{\lbrace {\vec{k}}, {\vec{q}} \rbrace}(0,0) \, .
\ee
We should mention that, unlike in the case of a 2d CFT such an expansion is not analytic (see (\ref{GCAOPE}) below).
The form of the two and three point function clearly exhibit essential singularities. Nevertheless
we will go ahead with the expansion assuming it makes sense in individual segments such as $x, t>0$. 
One can find the first few coefficients $C_{12}^{p\lbrace  {\vec{k}},
{\vec{q}} \rbrace}(t,x)$ by considering the three point 
function of the primary fields $\langle \phi_3 \phi_1 \phi_2 \rangle$ 
in the situation where $\phi_1$ approaches $\phi_2$.
In such a situation one can replace $\phi_1 \phi_2$,
in the three point function, with the RHS of \eq{ope}
and obtain
\be{opecor}
\langle \phi_3(t',x') \phi_1(t,x)\phi_2(0,0) \rangle  
=
\sum_{p, \lbrace {\vec{k}} , {\vec{q}} \rbrace}   
C_{12}^{p\lbrace {\vec{k}} , {\vec{q}} \rbrace}
(t,x) \langle
\phi_3(t',x')
\phi_p^{\lbrace  {\vec{k}}, {\vec{q}} \rbrace}(0,0) 
\rangle. 
\ee

We can find 
$C_{12}^{p\lbrace  0,0 \rbrace}$,
$C_{12}^{p\lbrace  1,0 \rbrace}$ and 
$C_{12}^{p\lbrace  0,1 \rbrace}$ by 
expanding the left hand side (LHS) of \eq{opecor}
with respect to the small parameter
$t \over t'$ while keeping 
$x' \over t'$ and $x \over t$ finite,
and comparing the $(t',x')$-dependence of the
both sides.
To make the final formulae simple, we concentrate on the case with
$\Delta_1=\Delta_2=\Delta$ and  $\xi_1=\xi_2=\xi $.

The expansion of the LHS is given as
\begin{align}
&\langle \phi_3(t',x') \phi_1(t,x)\phi_2(0,0) \rangle \notag \\
& \, = 
C_{312} 
t^{-2 \Delta + \Delta_3} 
{t'}^{-\Delta_3}
(t'-t)^{-\Delta_3}
\exp[ \xi_3 {x' - x \over t'-t }
+(2 \xi - \xi_3){x \over t}
+\xi_3 {x' \over t'}] \notag\\
& \, = C_{312} {t'}^{-2\D_3} {\rm e}^{2 \xi_3
{ x' \over t'}} \cdot t^{-2\D + \D_3} 
{\rm e}^{(2 \xi - \xi_3) {x \over t}} 
[1 + \{ \D_3 - \xi_3({x \over t} 
- {x' \over{t'}}) \} {t \over {t'}} + {\cal O}((t/t')^2) ],
\label{LHSexp}
\end{align}
while the RHS is given by
\ben{RHS}
&&\hspace{-1cm} \sum_{p, \lbrace {\vec{k}} , {\vec{q}} \rbrace}   
C_{12}^{p\lbrace {\vec{k}} , {\vec{q}} \rbrace}(t,x) 
\langle \phi_3(t',x') \phi_p^{\lbrace  {\vec{k}}, {\vec{q}}
\rbrace}(0,0) \rangle \nonumber \\  
&=& [C_{12}^{3\lbrace {0} , {0} \rbrace} (t,x) + C_{12}^{3\lbrace {1} ,
{0} \rbrace} (t,x) {\hat{L}}_{-1} + C_{12}^{3\lbrace {0} , {1} \rbrace}
(t,x) {\hat{M}}_{-1} + \ldots ] ({t'}^{-2\D_3} {\rm e}^{2 \xi_3 {x' \over{t'}}}) \nonumber \\
&=& {t'}^{-2\D_3} {\rm e}^{2 \xi_3 {x' \over{t'}}} 
[C_{12}^{3\lbrace {0} , {0}
\rbrace} + C_{12}^{3\lbrace {1} , {0} \rbrace} (2 \D_3 + 2 \xi_3 {x'
\over{t'}}) {t'}^{-1} + C_{12}^{3\lbrace {0} , {1} \rbrace} (2 \xi_3)
{t'}^{-1}+ \cdots ].
\een 
One can easily read off the coefficients by comparing \refb{LHSexp} and \refb{RHS} :
\ben{coef}
C_{12}^{3\lbrace {0} , {0} \rbrace} &=& C_{312} t^{-2\D + \D_3} 
{\rm e}^{(2 \xi - \xi_3) {x \over t}} \,,\nonumber \\
C_{12}^{3\lbrace {1} , {0} \rbrace} &=&  {1\over 2} C_{312} t^{-2\D +
\D_3+1} {\rm e}^{(2 \xi - \xi_3) {x \over t}}\,, \\
C_{12}^{3\lbrace {0} , {1} \rbrace} &=& - {1\over 2} C_{312} x t^{-2\D +
\D_3} {\rm e}^{(2 \xi - \xi_3) {x \over t}} \,.\nonumber
\een 
So, in this case we have the GCA OPE as:
\begin{align}
&\phi_1(t,x) \phi_2(0,0) \notag \\
& = \sum_{p} C_{p12} t^{-2\D + \D_p} 
{\rm e}^{(2 \xi - \xi_p) {x \over t}} 
\left( \phi_p(0,0) + {t\over 2} 
\phi_p^{\lbrace {1}, {0} \rbrace}(0,0) 
- {x\over 2} \phi_p^{\lbrace  {0}, {1} \rbrace}(0,0) + 
\ldots \right) \,. 
\label{GCAOPE}
\end{align}


\section{Fusion Rule for $\phi_{1(2,2)}$}

Here we consider the level two null state corresponding to the primary
\be{fus2}
\Delta = \Delta_{1(2,2)} =  -1\,, \quad \xi =\xi_{1(2,2)} = 0 \,,
\ee
and study the fusion rule for the
product $[\phi_{\Delta,\xi}] \times [\phi_{\Delta_1 , \xi_1}]$,
where $\Delta_1, \xi_1$ are given by \eq{fus1-2} and \eq{fus1-3}
with a generic triple $r (s,s')$. 

From the GCA in this case we have level two null states (see below \eq{null2}) given by
$|\chi^{(1)} \rangle = M_{-1}^2|\Delta, 0\rangle$ and $|\chi^{(2)} \rangle =  L_{-1}M_{-1}|\Delta=-1, 0 \rangle$. However, these null states are actually just descendants of the level one null state $M_{-1}|\Delta=-1, 0 \rangle$. The imposition of the GCA constraint
\ben{gca2cons}
&&{\hat M}_{-1}\langle \phi_{\Delta,\xi} (t, x)
\phi_{\Delta_1,\xi_1}(t_1,x_1) 
\phi_{\Delta_2, \xi_2}(t_2, x_2) 
\rangle  
\nonumber\\
&&\qquad = - \partial_x\langle \phi_{\Delta,\xi} (t,x)
\phi_{\Delta_1,\xi_1}(t_1,x_1) \phi_{\Delta_2,\xi_2}(t_2,x_2) \rangle = 0\,,
\een
simply gives $\xi_1=\xi_2$. Examination of \eq{fus1-2} and \eq{fus1-3}
immediately implies that the $r$ value of 
$\phi_{\Delta_2,\xi_2}(t_2,x_2)$ is the same as 
that of $\phi_{\Delta_1,\xi_1}(t_1,x_1)$ but its
$(s,s^{\prime})$ values are undetermined. 
As we will see below, in taking the nonrelativistic limit of the 
2d CFT we can extract the fusion rule obeyed in the relativistic CFT. 

Thus, 
let us  take the nonrelativistic limit of the relativistic fusion rule involving the primary $\phi_{2(1,1)}$.  
The conformal weights corresponding to $2(1,1)$
and $r(s,s')$ are expanded as
\ben{fus3}
h&=&h_{1,2} = -\frac{1}{2}+{3\epsilon \over 4C_2} + \O(\epsilon^2)\,, \quad \bar{h} =\bar{h}_{1,2} = -\frac{1}{2} +{3\epsilon \over 4C_2} +\O(\epsilon^2)\,;\nonumber \\
h_1&=& h_{r,s} = \frac{C_2}{4 \epsilon} (r^2 -1) + 
( {C_1 \over 4}(1-r^2) + {13\over 24} r^2-{r s\over 2} -
{1 \over 24}) -
\frac{ \epsilon}{4C_2} (r^2-s^2)+\O(\epsilon^2)\,,\nonumber \\
\bar{h}_1&=&\bar{h}_{r,s^{\prime}}= \frac{C_2}{4 \epsilon} (1- r^2) +
( {C_1 \over 4}(1-r^2) + {13 \over 24} r^2-
{r s^{\prime} \over 2} 
- {1 \over 24})
+\frac{\epsilon}{4 C_2} (r^2-{s^{\prime}}^2) + \O(\epsilon^2)\,,\nonumber
\een
respectively. 
Here in addition to the terms in \eq{hh1} and \eq{hh2},
if we maintain the terms of order ${\cal O}(\epsilon)$
then we can extract the information in the relativistic fusion rule.

The differential equation for the three point function
$\langle \phi_{1(2,2)} \phi_1 \phi_2 \rangle$ now 
implies that the $h$'s 
obey the equation
\be{fus4}
\frac{3}{2} (h_2-h-h_1) (h_2-h-h_1-1) 
+ (h_2 - h - 2h_1) (2h+1) =0\,,
\ee
and the same for $\bar h$'s.
Solving these equations, keeping the ${\cal O}(\epsilon)$ terms, we get the solutions as
\be{hoge}
h_2 = h_{r, s \pm 1} \,, \quad \bar{h}_2 
= \bar{h}_{r,s^{\prime} \pm 1 } \,,\nonumber
\ee
which tell us that we have four possible families 
appearing in the OPE between $[\phi_{1(2,2)}]$ and $[\phi_{r(s,s^{\prime})}]$.
Note that the $(s,s')$ values are now constrained.
In terms of the GCA weights, the four families are
specified as  
\be{res2}
\Delta_2 = \Delta_{r(s\pm1,s^{\prime} \pm 1)} \,, \qquad\xi_2= \xi_{r
(s \pm 1,s^{\prime} \pm 1)}\,,
\ee
where both the $\pm$ signs are taken independently.
 



\section{Asymptotic Form of the Hypergeometric Function} 
\label{AsymHyper}
\subsection{Saddle Point Analysis}
\label{saddle}
In this section we perform the saddle point analysis
of the integral formula for the Hypergeometric function:
\begin{equation}
F(\alpha,\beta,\gamma,z)
=
{\Gamma(\gamma) \over \Gamma(\beta) \Gamma(\gamma-\beta)}
\int_0^1 dw 
w^{\beta - 1}
(1 - w)^{\gamma - \beta -1}
(1 - zw)^{-\alpha}. \label{AppIntHyper}
\end{equation}
This formula is valid for 
the parameter ${\rm Re}(\gamma) > {\rm Re}(\beta) > 0$,
and we will use it for ${\rm Re}(z)<0$.
The parameters $\alpha$, $\beta$ and $\gamma$ are related 
to the physical parameters as 
\begin{eqnarray}
&& \alpha + \beta + 1 = 2 (\beta_1 + \beta_2 + \kappa), \quad
\gamma = 2 \beta_1 + \kappa, \\ 
&& \alpha\beta
=
\beta_1(\beta_1 - 1) 
+ 2\beta_1 \beta_2 
+ \beta_2(\beta_2 - 1) 
+ \kappa (2 \beta_1 + 2 \beta_2 + h_0 - h_3)\,,
\end{eqnarray}
with $\kappa = (2/3)(2h_0 + 1)$
and $\beta_i$ defined as solutions of the equations
\begin{equation}
\beta_i(\beta_i - 1 ) + \kappa \beta_i - \kappa h_i = 0,
\quad (i = 1,2).
\end{equation}
We expand parameters as 
\begin{equation}
\beta_i 
= {1 \over \epsilon} \beta_i^{(-1)}
+ \beta_i^{(0)} + \cdots, \quad
\alpha
= {1 \over \epsilon} \alpha^{(-1)}
+ \alpha^{(0)} + \cdots,
\end{equation}
etc.\,.
Then the quadratic equations for $\beta_i$
are solved as 
\begin{equation}
\beta_i^{(-1)} 
=
-{1 \over 2} C_2 (1 + \delta_i r_i), \quad
\beta_i^{(0)}
=
-{\delta_i \over 2} \ell_i - {1 \over 2} (\kappa^{(0)} - 1),
\end{equation}
with
\begin{eqnarray}
\ell_i 
& = &
{(\kappa^{(0)}-1)\kappa^{(-1)} + 2 \kappa^{(-1)} h_i^{(0)} + 2
\kappa^{(0)}h_i^{(-1)} \over C_2 r_i} \notag \\
& = &
{1 \over r_i}
\bigg\{
{1 \over 3} \Delta_0 
- {2 \over 3} 
+ 2 h_i^{(0)}
+ {1 \over 3} \Delta_0 r_i^2
+ {1 \over 3} r_i^2
\bigg\}. \label{Appli}
\end{eqnarray}
Since $\beta_i$ are auxiliary parameters
introduced in order to change the form of 
the differential equation 
(see the explanation around (\ref{beta_i})), 
we take 
$(\delta_1,\delta_2) = \delta_{C_2}(-1,+1)$
without loss of generality,
where $\delta_{C_2}$ is the sign of $C_2$.
Also, since the Hypergeometric function is symmetric 
under the exchange $\alpha \leftrightarrow \beta$,
we take $\alpha^{(-1)} < \beta^{(-1)}$, 
and then $\alpha$, $\beta$ and $\gamma$
are expanded as 
\begin{equation}
\begin{array}{lll}
\displaystyle
\alpha^{(-1)}
=
{|C_2| \over 2}(r_1 - r_2 - r_3), \,\,&
\displaystyle \beta^{(-1)}
= 
{|C_2| \over 2} (r_1 - r_2 + r_3), \,\, &
\displaystyle
 \gamma^{(-1)} = |C_2| r_1,\\[3mm]
\displaystyle 
\alpha^{(0)} 
= {\delta_{C_2}\over 2}(\ell_1 - \ell_2 - \ell_3) + {1 \over 2}, \,\, & 
\displaystyle
\beta^{(0)} =
{\delta_{C_2} \over 2 } (\ell_1 - \ell_2 + \ell_3) + {1 \over 2}, \,\, & 
\gamma^{(0)} = \delta_{C_2} \ell_1 + 1.
\end{array} \label{app_parameters}
\end{equation}
Here $\ell_3$ is defined in the same way as $(\ref{Appli})$.

Now the integral (\ref{AppIntHyper}) is written as 
\begin{equation}
F(\alpha,\beta,\gamma,z)
=
{\Gamma(\gamma) \over \Gamma(\beta) \Gamma(\gamma - \beta)}
\int_0^1 dw 
{\rm e}^{{1 \over \epsilon} 
g^{(-1)}(w,t) +
x \partial_t g^{(-1)}(w,t) 
+
g^{(0)}(w,t)
+
{\cal O}(\epsilon)}, \label{AppIntHyperExp}
\end{equation}
where $g^{(n)}(w,z)$'s are defined as 
the functions appearing in the expansion of the 
exponent of the integrand as
\begin{equation}
(\beta-1) \log w 
+
(\gamma - \beta -1)\log(1-w)
-
\alpha \log (1-zw)
=
\sum_{n} \epsilon^n g^{(n)}(w,z).
\end{equation}
In (\ref{AppIntHyperExp}), we have further 
expanded based on $z = t + \epsilon x$.
The saddle points of the integrand are 
given by $\partial_w g^{(-1)}(w_\ast,t)=0$
and they are located at
\begin{equation}
w_\ast^\pm
=
{
\gamma^{(-1)} - (\alpha^{(-1)} - \beta^{(-1)})t \mp |C_2|\sqrt{D(t)}
\over 
2
(\gamma^{(-1)} - \alpha^{(-1)})t
}.
\end{equation}
By using the fact that $r_i$'s satisfy the 
triangle inequality, which follows from the
fusion rule (see the next section), we can show that 
both the saddle points are on the real axis,
namely $D(t) \geq 0$.
In fact, for the parameterization explained above,
only one ($w = w_\ast^+$) of the saddle points 
is located on the segment $0<w<1$, and the integral
is evaluated by taking it.
By taking the saddle point value and 
doing the Gaussian integral, we obtain 
the following asymptotic form of the 
Hypergeometric function for small $\epsilon$:
\begin{equation}
F(\alpha,\beta,\gamma,z)
\to
\Big({\epsilon \over 2 \pi} \Big)^{1 \over 2}
{\Gamma(\gamma) \Gamma(1 - \alpha) \over \Gamma(\gamma - \alpha)}
{\cal K}_+(t,x),
\quad (t<1), \label{app_Hyper_asymp}
\end{equation}
where the function ${\cal K}_+$ is defined by
\begin{eqnarray}
{\cal K}_+ (t,x)
&=&
|C_2|^{-{1 \over 2}}
\big(D(t)\big)^{-{1 \over 4}} 
\bigg\{
{r_1 + r_2 + r_3 \over r_1 + r_2 - r_3}
{-r_1 + r_3 t + \sqrt{D(t)} \over r_1 + r_3 t+ \sqrt{D(t)}}
{1 \over t}
\bigg\}^{{1 \over \epsilon} {|C_2| \over 2}r_1 + {\delta_{C_2}\over 2} \ell_1} \notag
\\
&& \times \bigg\{
{r_1 + r_2 + r_3 \over r_1 + r_2 - r_3}
{r_2 - r_3 (1-t) + \sqrt{D(t)} \over r_2 + r_3 (1-t) - \sqrt{D(t)}}
(1-t)
\bigg\}^{{1 \over \epsilon} {|C_2| \over 2}r_2 + {\delta_{C_2}\over 2} \ell_2}
\notag \\ && \times 
\bigg\{
{r_1 + r_2 - r_3 \over - r_1 + r_2 + r_3}
{r_2 + r_3 (1-t) + \sqrt{D(t)} \over r_2 - r_3 (1-t) + \sqrt{D(t)}}
\bigg\}^{{1 \over \epsilon} {|C_2| \over 2}r_3 + {\delta_{C_2} \over 2} \ell_3}
\notag \\
&& \times 
\exp\bigg\{
{x |C_2| \over2 t (1-t)} \big( - r_1(1-t) - r_2 t + \sqrt{D(t)} \big)
\bigg\} \label{calK(1)} \\
& = & K_+(t)  
\exp\bigg\{
{x |C_2| \over 2t (1-t)} 
\big( - r_1(1-t) - r_2 t + \sqrt{D(t)} \big)
\bigg\}.
\label{calK(2)}
\end{eqnarray}
The function $K_+(t) = {\cal K}_+(t,0)$ 
is introduced for convenience.
The $x$-dependence appears only in the 
exponential form, which can be derived by expanding 
$K_+ (z = t + \epsilon x)$ 
with respect to $\epsilon$, 
as it can be seen from (\ref{AppIntHyperExp}).
The overall factor $\epsilon^{1 \over 2}$
on the RHS of (\ref{app_Hyper_asymp})
is cancelled by the factor $\epsilon^{-{1 \over 2}}$
when we expand the Gamma functions.

\subsection{Fusion Rule and Relation Between $r_i(s_i,s_i')$'s}
\label{FusionRuleAnd}
In the main text and this appendix, relations between three 
primary fields $\phi_1$, $\phi_2$ and $\phi_3$
in the four point function 
$\langle \phi_0\phi_1\phi_2\phi_3 \rangle$
with $\phi_0 = \phi_{2(1,1)}$
are used in several places.
These relations are derived from the 
generic fusion rule in the relativistic 2d CFT:
\begin{equation}
[\phi_{r_i(s_i,s'_i)}] 
\times 
[\phi_{r_j(s_j,s'_j)}]
=
\sum_{r_p =|r_i - r_j| + 1}^{r_i+r_j-1}
\sum_{s_p =|s_i - s_j| + 1}^{s_i+s_j-1}
\sum_{s_p' =|s'_i - s'_j| + 1}^{s'_i+s'_j-1}
[\phi_{r_p(s_p,s'_p)}].
\end{equation} 
The variables $r_p$, $s_p$ and $s_p'$
are incremented by $2$.
Let us discuss only the channel 
in which we take the OPE between
$\phi_0$ and $\phi_1$ first,
then we take the OPE between
the intermediate field $\phi_p$ 
and $\phi_2$ to give the remaining
field $\phi_3$.
Symbolically, we consider the order
$\langle \phi_0 \phi_1 \phi_2 \phi_3 \rangle 
=\Sigma_p C_{01p}\langle \phi_p \phi_2 \phi_3 \rangle
=\Sigma_{p,q} C_{01p} C_{p2q} \langle \phi_q \phi_3 \rangle $.
Here in the last step, because of the orthogonality 
of the conformal families, we have $r_q(s_q,s_q') = r_3(s_3,s_3')$.
The other cases can be discussed similarly.

In this case, the possible intermediate state is given by 
$r_p(s_p,s_p') = (r_1\pm1) (s_1,s_1')$ and the fusion
rule for the OPE between $\phi_p$ and $\phi_2$ is 
given by
\begin{align}
r_3 & = |r_1\pm1 - r_2|+1,\,\, |r_1\pm 1 - r_2|+3,  \,\, \cdots, \,\,
 r_1\pm 1+r_2 - 1\,, 
\label{r-fusion} \\
s_3 & = |s_1- s_2|+1, \,\, |s_1 - s_2| + 3, \,\, \cdots, \,\, s_1+s_2-1\,, 
\label{s-fusion} \\
s_3' & = |s_1' - s_2'|+1, \,\, |s_1' - s_2'| + 3, \,\, \cdots,\,\, s_1'+s_2'-1 \,.
\label{s'-fusion}
\end{align}

From the first equation (\ref{r-fusion}),
we see that the three $r_i$'s 
satisfy the triangle inequality, 
which is important in the saddle point analysis in appendix \ref{saddle}.
Also we notice that the possible values 
for the summation of three $r_i$'s are even, 
while the summation of three $s_i$'s and $s_i'$'s
are always odd numbers. 
It is consistent with the property that 
$s_i' - s_i$ are even integers (see Sec. \ref{Sec:Singlevalued}).

\subsection{Validity of the Analysis}
Let us now clarify the validity of the analysis in 
appendix \ref{saddle}.
As we mentioned, the integral formula
(\ref{AppIntHyper}) is valid for the parameter
satisfying $\gamma >\beta > 0$.
From the expansion (\ref{app_parameters})
with (\ref{Appli}), it is easy to see that 
for the case with $r_1-r_2+r_3 = 0$ or 
$r_1 + r_2 - r_3 = 0$, namely, $\beta_{-1}=0$
or $\gamma_{-1} = \beta_{-1}$,
the condition is violated for general $s_i$'s.
Hence for these cases, the simple integral 
formula (\ref{AppIntHyper}) is not available
and we need to consider more generic formula,
which we do not pursuit in the present paper.
Examples of such situations include the 
case when one of the fields $\phi_{1,2,3}$
is the identity operator.
On the other hand, in the case with 
$r_1 - r_2 - r_3 = 0$, 
the above analysis itself is valid.
However, for deriving the asymptotic form 
for the rest of the region, namely, $t > 1$ and 
also for the other basis function in the next subsection,
the case $r_1 - r_2 - r_3 = 0$
violates the conditions which 
are equivalent to the ones explained above.
Hence our derivation of the asymptotic form 
of the four point function is applicable only 
for the cases with $r_1<r_2+r_3$, $r_2<r_1+r_3$ and 
$r_3<r_1+r_2$, i.e. the strict triangle inequality
is satisfied.

The validity of the saddle point analysis also 
requires the second derivative 
\begin{equation}
\partial_w^2 g^{(-1)}(w_\ast^+,t)
=
-{|C_2|\sqrt{D} \over (1 - t w_\ast^+)^2}
{
(2 \alpha_{-1} \beta_{-1} - \gamma_{-1}(\alpha_{-1} + \beta_{-1}))t
+
\gamma_{-1}^2 + \gamma_{-1} |C_2| \sqrt{D}
\over 
2 \beta_{-1} (\gamma_{-1} - \beta_{-1})
},
\end{equation}
to be negative definite (with large $1/\epsilon$ in front of it).
This condition is perfectly satisfied with the above assumption
for the parameters.

\subsection{A Complete Set of the Solutions and Useful Formulae}

Next we derive the other solution 
$(-z)^{1-\gamma} F(\alpha-\gamma+1,\beta-\gamma+1,2-\gamma,z)$
for the region $t<0$
by using (\ref{app_Hyper_asymp})
and the formula (\ref{FtoF+F(1)}) (formula (I)).
By changing the parameters of the formula (I)
as 
$\alpha \to \alpha - \gamma +1$,
$\beta \to \beta - \gamma +1$
and $\gamma \to 2 - \gamma$,
and multiplying the factor $(-z)^{1-\gamma}$
to the both sides of the resulting equation,
we obtain the similar formula (formula (I\!I)) for 
$(-z)^{1-\gamma} F(\alpha-\gamma+1,\beta - \gamma+1,2 - \gamma,z)$
expressed in terms of the same two functions 
on the RHS of (\ref{FtoF+F(1)}).
Regarding the second function on the RHS, i.e.,
$(-1)^{-\beta}F(\beta,\beta-\gamma+1,\beta-\alpha+1,1/z)$,
the condition $\alpha < 0 < \beta < \gamma$, which 
we are now assuming to hold, 
assures $\alpha' < 0 < \beta' < \gamma'$ with 
$\alpha' = \beta - \gamma + 1$, 
$\beta' = \beta$ and $\gamma' = \beta - \alpha + 1$.
Then the asymptotic form (\ref{app_Hyper_asymp}) is 
applicable for this case with the parameter 
change $\alpha \to \alpha'$, 
$\beta \to \beta'$
and $\gamma \to \gamma'$ along with the 
coordinate change $z \to 1/z$. 
Note that the condition $\alpha^{(-1)} = {|C_2|\over 2}(r_1-r_2 -r_3)<  0$
is necessary here in order for the condition 
$\beta' < \gamma'$ to be satisfied in general,
as we mentioned in the previous subsection.

Then by using the two formulae (I) and (I\!I), 
which are mentioned above, we can write 
down the asymptotic form of the 
function 
$(-z)^{1-\gamma} F(\alpha-\gamma+1,\beta - \gamma+1,2 - \gamma,z)$
in terms of the asymptotic forms 
of the functions $F(\alpha,\beta,\gamma,z)$
and $(-z)^{-\beta} F(\alpha',\beta',\gamma',1/z)$.
The explicit form is given by
\begin{eqnarray}
&&(-z)^{1-\gamma}
F(\alpha-\gamma+1, \beta - \gamma + 1, 2 - \gamma,z) 
 \to 
\Big(
{\epsilon \over 2 \pi}
\Big)^{1 \over 2}
\bigg\{
{\Gamma(\beta) \Gamma(2 - \gamma) \over \Gamma(\beta - \gamma + 1)}
K_+(z) \notag \\
&&\qquad \qquad  +
{\Gamma(\beta) \Gamma(\gamma-\beta) \over \Gamma(\gamma - 1)}
(-z)^{-{1 \over \epsilon} |C_2| r_1 - \delta_{C_2}\ell_1}
(1-z)^{{1 \over \epsilon} |C_2| r_2 + \delta_{C_2}\ell_2}
K_-(z)
\bigg\}.
\label{zF}
\end{eqnarray}
Here again $z=t + \epsilon x$ with
small $\epsilon$ is assumed.
The function $K_-(z)$ 
can be obtained from
$K_+(z)$ by replacing as 
$1 \leftrightarrow 2$ and 
$(t,x) \leftrightarrow (1-t,-x) $,
or equivalently $z \leftrightarrow 1 -z$.

The asymptotic form of the anti-holomorphic functions
for the region $t<0$ can be derived in the same way. 
In our choice of parameters
(see the explanation after (\ref{abcexpansion})
for the parameter choice of the anti-holomorphic part), 
the parameters 
in the anti-holomorphic part satisfy
$\bar \gamma < \bar \beta < 0  < \bar \alpha$.
By finding appropriate basis functions
to which the asymptotic form (\ref{app_Hyper_asymp})
is applicable, and also by using the formula 
(\ref{FtoF+F(1)}), we obtain the following result:
\begin{align}
& F(\bar \alpha, \bar \beta, \bar \gamma,\bar z)
\to 
(2 \pi \epsilon)^{1 \over 2}
\bigg\{
{\Gamma(\bar \alpha - \bar \gamma + 1 )
\over
\Gamma(1 - \bar \gamma) \Gamma(\bar \alpha)}
\bar K_+(\bar z)
\notag \\
&\,\,\,+
{
\Gamma(\bar \alpha - \bar \gamma + 1) 
\Gamma(\bar \gamma)
\over
\Gamma(\bar \alpha) 
\Gamma(\bar \gamma - \bar \beta)
\Gamma(\bar \beta - \bar \gamma + 1)
}
(- \bar z)^{{1 \over \epsilon} |C_2| r_1 - \delta_{C_2} \bar \ell_1}
(1- \bar z)^{- {1 \over \epsilon} 
|C_2| r_2 + \delta_{C_2} \bar \ell_2}
\bar K_-(\bar z)
\bigg\}, \label{bF}\\[4mm]
& (- \bar z)^{1 - \bar \gamma}
F(\bar \alpha - \bar \gamma + 1, \bar \beta - \bar \gamma +
1,2-\gamma,\bar z) \notag \\
&\,\,\, \to 
(2 \pi \epsilon)^{1 \over 2}
{\Gamma(2 - \bar \gamma) 
\over 
\Gamma(1 - \bar \beta) \Gamma(\bar \beta -\bar \gamma + 1)}
(- \bar z)^{{1 \over \epsilon}|C_2|  r_1 - \delta_{C_2} \bar \ell_1}
(1- \bar z)^{- {1 \over \epsilon} |C_2|
 r_2 +\delta_{C_2} \bar \ell_2}
\bar K_-(\bar z), \label{bzF}
\end{align}
where $\bar K_\pm(\bar z)$ are 
obtained from $K_\pm(z)$ with replacements
$\epsilon \to - \epsilon$ and $\ell_i \to \bar \ell_i$.
In the main text, we use Eqs.(\ref{zF}), (\ref{bF}) and (\ref{bzF}) 
for deriving the asymptotic form of the four point function
in the region $t<0$. 

In fact, by means of (\ref{app_Hyper_asymp}),
we can derive the asymptotic 
forms of the sets of two independent solutions 
for the remaining segments.
Here is the list of the asymptotic forms:
\begin{enumerate}
\item 
The functions whose asymptotic forms are given in terms of $K_+(z)$:
\begin{itemize}
\item $t<1$
\begin{equation}
F(\alpha,\beta,\gamma,z)\to
\Big({\epsilon \over 2 \pi}\Big)^{1 \over 2} 
{\Gamma(\gamma) \Gamma(1-\alpha) \over \Gamma(\gamma - \alpha)}K_+(z)
\label{asymp(1)}
\end{equation}
\item $1<t$
\begin{equation}
z^{-\beta} F(\beta - \gamma + 1 , \beta, \beta - \alpha + 1, 
 1/z) \to
\Big({\epsilon \over 2 \pi}\Big)^{1 \over 2}
{\Gamma(\beta-\alpha+1)\Gamma(\gamma - \beta)\over \Gamma(\gamma -
\alpha)}
K_+(z) 
\label{asymp(2)}
\end{equation}
\end{itemize}
\item
The functions whose asymptotic forms are given in terms of $K_-(z)$:
\begin{itemize}
\item $t<0$
\begin{align}
&\hspace{-0.5cm}(-z)^{-\beta} F(\beta - \gamma + 1, \beta, \beta-\alpha+1,1/z)
\notag \\
&\hspace{-0.5cm}\to
\Big({\epsilon \over 2 \pi}\Big)^{1 \over 2}
{\Gamma(\beta - \alpha +1) \Gamma(\gamma - \beta) 
\over \Gamma(\gamma-\alpha)}
(-z)^{-{1 \over \epsilon}|C_2| r_1 - \delta_{C_2}\ell_1}
(1-z)^{{1 \over \epsilon}|C_2| r_2 + \delta_{C_2}\ell_2}
K_-(z) 
\label{asymp(3)}
\end{align}
\item $0<t$
\begin{align}
&\hspace{-0.5cm} z^{1-\gamma}(1-z)^{\gamma - \alpha - \beta}
F(1-\beta,1-\alpha,\gamma - \alpha - \beta +1,1-z) \notag \\
&\hspace{-0.5cm}\to
\Big({\epsilon \over 2 \pi}\Big)^{1 \over 2}
{\Gamma(\gamma - \alpha - \beta + 1) \Gamma(\beta) \over
\Gamma(\gamma - \alpha)}
z^{-{1 \over \epsilon} |C_2| r_1 - \delta_{C_2} \ell_1}
(1-z)^{{1 \over \epsilon}|C_2| r_2 + \delta_{C_2} \ell_2} K_-(z) 
\label{asymp(4)}
\end{align}
\end{itemize}
\item
The functions whose asymptotic forms are given in terms of 
$\bar K_+(\bar z)$:
\begin{itemize}
\item $t<0$
\begin{align}
&\hspace{-2cm}(- \bar z)^{\bar \beta - \bar \gamma}
(1 - \bar z)^{\bar \gamma - \bar \alpha - \bar \beta}
F(\bar \gamma -\bar \beta, 1 - \bar \beta,
\bar \alpha - \bar \beta + 1, 1/\bar z) \notag \\
& \hspace{3cm} \to
(2 \pi \epsilon)^{1 \over 2}
{
\Gamma(\bar \alpha - \bar \beta +1) \over \Gamma(1 - \bar \beta) 
\Gamma(\bar \alpha)}
\bar K_+(\bar z) 
\label{asymp(5)}
\end{align}
\item $0<t$
\begin{equation}
 F(\bar \beta ,\bar \alpha, \bar \alpha + \bar \beta - \bar \gamma + 1,
1-\bar z) 
\to
(2 \pi \epsilon)^{1 \over 2}
{\Gamma(\bar \alpha + \bar \beta - \bar \gamma + 1)
\over
\Gamma(\bar \alpha) \Gamma(\bar \beta - \bar \gamma + 1)}
\bar K_+(\bar z) 
\label{asymp(6)}
\end{equation}
\end{itemize}
\item
The functions whose asymptotic forms are given in terms of 
$\bar  K_-(\bar z)$:
\begin{itemize}
\item $t<1$
\begin{align}
&\hspace{-0.5cm}\bar z^{1-\bar \gamma} 
(1-\bar z)^{\bar \gamma - \bar \alpha- \bar \beta}
F(1 - \bar \alpha, 1 - \bar \beta, 2 - \bar \gamma, \bar z) \notag \\
&\hspace{-0.5cm}\to (2 \pi \epsilon)^{1 \over 2}
{\Gamma(2-\bar \gamma) \over \Gamma(1 - \bar \beta) \Gamma(\bar \beta -
\bar \gamma + 1)}
\bar z^{{1 \over \epsilon}|C_2| r_1 - \delta_{C_2} \bar \ell_1}
(1 - \bar z)^{-{1 \over \epsilon}|C_2| r_2 + \delta_{C_2} \bar \ell_2}
\bar K_-(\bar z) 
\label{asymp(7)}
\end{align}
\item $1<t$
\begin{align}
&\bar z^{\bar \beta - \bar \gamma}
(1-\bar z)^{\bar \gamma - \bar \alpha - \bar \beta}
F(\bar \gamma -\bar \beta,1 - \bar \beta, \bar \alpha - \bar \beta +1
,1/\bar z) \notag \\
& \to
(2 \pi \epsilon)^{1 \over 2}
{\Gamma(\bar \alpha - \bar \beta +1)
\over \Gamma(\bar \alpha)\Gamma(1- \bar \beta)}
\bar z^{{1 \over \epsilon} |C_2|r_1 - \delta_{C_2} \bar \ell_1}
(1-\bar z)^{-{1 \over \epsilon} |C_2| r_2 + \delta_{C_2} \bar \ell_2} 
\bar K_-(\bar z) 
\label{asymp(8)}
\end{align}
\end{itemize}
\end{enumerate}
The functions on the LHS's are the
solutions of the same Hypergeometric differential equation
(\ref{DiffHyper}). 
This basis
covers all the segments $t<0$, $0<t<1$ and $t>1$.
Note that these functions satisfy the 
parameter condition which 
is required for the application
of (\ref{app_Hyper_asymp}).
On the RHS's, 
$z=t+ \epsilon x$ and $\bar z = t - \epsilon x$
with the small parameter $\epsilon$ 
are assumed.
Further expansion with respect to this $\epsilon$
gives the $x$-dependent exponential factors.

Finally, we summarize some useful formulae (see for example \cite{formula}):
\begin{align}
 F(\alpha,\beta,\gamma,z) 
& = {\Gamma(\beta - \alpha)\Gamma(\gamma) \over \Gamma(\beta) \Gamma(\gamma -
 \alpha)}
(-z)^{-\alpha} F(\alpha,\alpha- \gamma + 1, \alpha - \beta + 1, 1/z )
 \notag \\
&  \quad + 
{\Gamma(\alpha - \beta) \Gamma(\gamma)
\over \Gamma(\alpha )\Gamma(\gamma - \beta)}
(-z)^{-\beta} F(\beta,\beta - \gamma+1, \beta - \alpha + 1, 1/z),
\label{FtoF+F(1)}\\
& =
{\Gamma(\alpha + \beta - \gamma) \Gamma(\gamma)
\over \Gamma(\alpha)\Gamma(\beta)}
(1-z)^{\gamma - \alpha - \beta}
F(\gamma-\alpha,\gamma-\beta,\gamma - \alpha- \beta + 1,1-z) \notag \\
& \quad +
{\Gamma(\gamma - \alpha - \beta) \Gamma(\gamma)
\over \Gamma(\gamma - \alpha) \Gamma(\gamma - \beta)}
F(\alpha,\beta,\alpha + \beta - \gamma + 1, 1 -z ). \quad
\label{FtoF+F(2)}
\end{align}
The functions appearing in (\ref{FtoF+F(1)}) and 
(\ref{FtoF+F(2)}) satisfy the Hypergeometric differential equation
independently.
The first form (\ref{FtoF+F(1)}) is valid 
for any $z$ except for the positive real number, 
while the second (\ref{FtoF+F(2)}) is valid 
for any $z$ except on the segment $[0,1]$ of the real axis.
By using the following formula:
\begin{align}
F(\alpha,\beta,\gamma,z) 
& = 
(1-z)^{\gamma - \alpha - \beta} F(\gamma - \alpha, \gamma -\beta, \gamma,z),
\end{align}
the functions on the LHS of
(\ref{asymp(1)})--(\ref{asymp(8)}) can be converted into 
the functions appearing in (\ref{FtoF+F(1)}) and
(\ref{FtoF+F(2)}) or 
$z^{1-z} F(\alpha -\gamma + 1, \beta - \gamma + 1, 2 - \gamma,z)$.


\end{document}